\documentclass[fleqn,usenatbib]{mnras}

\usepackage{newtxtext,newtxmath}

\usepackage[T1]{fontenc}

\usepackage{ulem}

\DeclareRobustCommand{\VAN}[3]{#2}
\let\VANthebibliography\thebibliography
\def\thebibliography{\DeclareRobustCommand{\VAN}[3]{##3}\VANthebibliography}

\usepackage{longtable}
\usepackage{natbib}
\usepackage{color}
\newcommand{\project}[1]{\textsl{#1}}

\newcommand{\apogee}{\project{\textsc{apogee}}}

\newcommand{\apokasc}{\project{\textsc{apokasc}}}

\newcommand{\Gaia}{\project{Gaia}}

\newcommand{\teff}{\mbox{$T_{\rm eff}$}}

\newcommand{\rgal}{\mbox{$\rm R_{gal}$}}

\newcommand{\logg}{\mbox{$\log g$}}

\usepackage{multirow}
\usepackage{multicol}

\usepackage{graphicx}	
\usepackage{amsmath}

\title[GalStar]{Milky Way Mapper decoded abundances -- I: Shared disc enrichment patterns}

\author[M. K. Ness et al.]{Melissa K. Ness$^{1}$\thanks{E-mail: melissa.ness@anu.edu.au},
Sarah Aquilina$^{1}$,
Jennifer Mead$^{2}$,
Emily Griffith$^{3}$,
Catherine Manea$^{4}$,
Jonathan Bird$^{5}$,
\newauthor
Andrew R. Casey$^{6,7}$,
Lucy (Yuxi) Lu$^{8,9}$,
Kathryn V. Johnston$^{2}$,
Michael R. Blanton$^{10}$,
James W. Johnson$^{11}$,
\newauthor
Maja Jablonska$^{1}$,
Leticia Carigi$^{12}$,
Jos\'e G. Fern\'andez-Trincado$^{13}$,
Ricardo L\'opez Valdivia$^{14}$,
Ying-Yi Song$^{15,16}$,
\newauthor
Juna Kollmeier$^{11,17,18}$, 
\\
$^{1}$Research School of Astronomy and Astrophysics, Australian National University, Canberra, ACT 2611, Australia\\
$^{2}$Department of Astronomy, Columbia University, Pupin Physics Laboratories, New York, NY 10027, USA\\
$^{3}$Center for Astrophysics and Space Astronomy, Department of Astrophysical and Planetary Sciences, University of Colorado, 389 UCB, Boulder, CO 80309-0389, USA\\
$^{4}$Department of Physics \& Astronomy, University of Utah, Salt Lake City, UT 84112, USA\\
$^{5}$Department of Physics and Astronomy, Vanderbilt University, 6301 Stevenson Center, Nashville, TN 37235, USA\\
$^{6}$Center for Computational Astrophysics, Flatiron Institute, 162 5th Ave, New York, NY 10010, USA\\
$^{7}$School of Physics and Astronomy, Monash University, VIC 3800, Australia\\
$^{8}$Department of Astronomy, The Ohio State University, 140 W 18th Ave, Columbus, OH 43210, USA\\
$^{9}$Center for Cosmology and Astroparticle Physics (CCAPP), The Ohio State University, 191 W. Woodruff Ave., Columbus, OH 43210, USA\\
$^{10}$Center for Cosmology and Particle Physics, Department of Physics, New York University, 726 Broadway Rm. 1005, New York, NY 10003, USA \\
$^{11}$Carnegie Science Observatories, 813 Santa Barbara Street, Pasadena, CA 91101, USA\\
$^{12}$Instituto de Astronomía, Universidad Nacional Aut\'onoma de M\'exico, A.P. 70-264, 04510 CDMX, México \\
$^{13}$Universidad Cat\'olica del Norte, N\'ucleo UCN en Arqueolog\'ia Gal\'actica - Instituto de Astronom\'ia, Av. Angamos 0610, Antofagasta, Chile\\
$^{14}$Universidad Nacional Aut\'onoma de M\'exico, Instituto de Astronom\'ia, AP 106, Ensenada 22800, BC, M\'exico \\
$^{15}$Dunlap Institute for Astronomy \& Astrophysics, University of Toronto, 50 St. George Street, Toronto, ON M5S 3H4, Canada\\
$^{16}${David A. Dunlap Department of Astronomy \& Astrophysics, University of Toronto, 50 St. George Street, Toronto, ON M5S 3H4, Canada} 
$^{17}$Canadian Institute for Theoretical Astrophysics, University of Toronto, Toronto, ON M5S 3H8, Canada\\
$^{18}$Canadian Institute for Advanced Research, 661 University Avenue, Suite 505, Toronto, ON M5G 1M1, Canada }

\date{Accepted YYYY Month DD. Received YYYY Month DD}

\pubyear{2024}

\begin{document}
\label{firstpage}
\maketitle

\begin{abstract}
Elemental abundances in the Milky Way disc trace its star-formation and enrichment history, but predicting these abundances from theory is limited by uncertain nucleosynthetic yields and poorly constrained chemical evolution models. Large surveys provide many abundances that enable multi-dimensional insight. However, having so much data available complicates joint visualisation and physical interpretation. Here, we examine the element abundances of 70,057 red giant stars from the Milky Way Mapper survey ([Fe/H]~$> -1$), using 16 elements (O,~Mg,~Al,~Si,~S,~K,~Ca,~Ti,~V, ~Cr, Mn,~Fe,~Co,~Ni,~Ce,~Nd). To tackle the challenges of joint-interpretation of these elements, we build a generative data-driven model, expressing each star’s abundance vector as a linear combination of a few ($4$) latent nucleosynthetic \textit{patterns}. These patterns are shared among the population but vary in \textit{fraction} between stars. The model accurately generates the measured abundances, with $\chi^2 < 3$~(5) for $\sim$~80\%~(95\%) of stars. Model failures, where stars' abundances are not generated by the latent basis reveal accreted material and the role of multiple channels of metal-poor disc enrichment. We associate the recovered patterns, which represent high-precision ($\sigma_P \sim 3$\%) nucleosynthetic channels, with specific enrichment sources; (early and late) core-collapse supernovae, supernovae Type Ia, and asymptotic giant branch stars. We subsequently explore how the dominance of enrichment channels varies across age, metallicity and spatial extent of the disc, and show that enrichment patterns tightly couple to orbital properties.  Mean pattern fractions vary smoothly with enrichment, and change rapidly across the valley between the high- and low-$\alpha$ sequences. Our results provide a framework for improving our understanding of Galactic evolution in the Milky Way.
\end{abstract}


\section{Introduction}

Stellar abundances are set by the evolutionary history of the star-forming environment. Although some elements, such as carbon and nitrogen, can change significantly during the lifetime of a star due to internal evolution, most are inherited from the star's birth environment and are considered unchanged thereafter \citep{Freeman2002}. Therefore, an individual star's elemental abundances reflect the integrated nucleosynthetic yields of previous stellar generations, modulated by the star formation history of the system up to the time of the star’s birth.  Subsequently, element abundances represent an evolutionary snapshot in the time of the Galaxy, or the `chemical fingerprint' of birth environment. 

Recent analyses of both large survey data and smaller high-fidelity samples have demonstrated that for stars across the Milky Way disc, a subset of elements (2-4) can predict up to 30 others, to within 5-10\% \citep{Mead2025, Griffith2024, Weinberg2019, Weinberg2022, Ness2022}.  In other words, the multi-element abundance space of disc stars is highly correlated and low-dimensional. Simple models conditioned on a small set of elemental abundances (or analogously, on age and metallicity) can reproduce the full abundance patterns of disc stars with high precision. The small residuals away from these predictive models express the information captured in each element that is not represented by the subset of dominant contributions.  Within these residuals are key signatures of the nucleosynthetic history and the nature of the underlying sources \citep{Griffith2025, Ting2021, Mead2025, Manea2025}. 

The strong correlation between elements in the disc opens up opportunities for connecting observations to theory. The inter-element correlation structure and residual amplitudes away from low-dimensional predictive models are key constraints on chemical evolution models. In addition, the ability to accurately generate the abundances from a lower-dimensional representation enables the construction of modeling frameworks to re-project individual abundance vectors into population statistics. 

In this work, we leverage this latter opportunity and take the first steps toward building a fully data-driven chemical evolution framework of the disc.  We want to access the signatures of the chemical evolution and signatures in the mode of star formation by connecting abundances to their sources and source enrichment over time. To do this, we use data from the SDSS V Milky Way Mapper Survey \citep{Kollmeier2025}.  We analyse the SDSS V stars in the DR19 release \citep{DR19, Meszaros2025}, selecting 70,057 stars from \rgal\ $= 0.5 - 20$~kpc with high-fidelity elemental abundances, to build a data-driven model of 16 chemical elements. Ages can be inferred for these stars via data-driven models that use the spectra or [C/N] ratios and train a model using subset of \apokasc\ stars \citep{Pins2025} labeled with asteroseismic ages \citep{Martig2016, Ness2016, Lu2022a, Leung2023, StoneM2025}. With spectroscopic ages, we have the capability to examine the abundance representations over time, and to decode the formation history of the disc. 

Our goal is to trace the changing contributions of nucleosynthetic sources across the Galaxy and over time, and to uncover the imprints of the star formation history on the chemical landscape of the disc. In Section \ref{data} we introduce the data used for our analysis. In Section \ref{method} we describe the mathematical formalism of the problem and the non-negative matrix factorisation (NMF) that we employ to learn the pattern vector and weighting (fractional contribution of the pattern) for each star and validate the model. In Section 4 we associate pattern vectors with nucleosynthetic sources, examine model failures, examine the distribution of fractional patterns across abundance, spatial, orbital and temporal extent. In  Section \ref{discussion} we discuss the implications of our findings as well as future prospects for latent space modeling. We conclude in Section \ref{conclusion}.

\section{Data}
\label{data}

We use the version-5 of the SDSS-V ASTRA results (Casey et al., in preparation)  \citep[DR19;][]{Kollmeier2025, DR19}. SDSS-V utilises the Sloan Foundation Telescope at Apache Point Observatory \citep{Gunn2006} and the du Pont Telescope at Las Campanas Observatory \citep{Bowen1973}, both equipped with APOGEE spectrographs \citep{Wilson2019}. 

We select red giant stars observed at resolution $R=22,500$ with the infrared \apogee\ spectrograph ($\approx1.51-1.7\mu m$) from the Milky Way Mapper survey.

\begin{equation}
\begin{aligned}
& H < 11 \;\;\text{AND} \\
& (
      \varpi < 100 \times 10^{0.2(-2.5 - H)}
      \;\;\text{OR}\;\;
      (G - H > 5)
      \;\;\text{OR}\;\;
      \text{Gaia non-detection})
\end{aligned}
\end{equation}

This selection includes is a mapping of the disc and bulge of the Milky Way and extends the revolutionary reach of the predecessor \apogee\ survey \citep{Majewski2017}.

We use the following selection criteria on the measurements of the spectra, from the version-0.5.0 ASTRA file `astraAllStarASPCAP-0.5.0'\footnote{For general survey characteristics and publicly released data products,
we refer to the SDSS Data Release documentation \citep[e.g.][]{DR19}.} with stellar parameters and abundances measured via the \textsc{ASTRA/ASPCAP} framework (Casey et al., in preparation; \citet{GP2016} and also see \citet{Nidever2015a, Holtzman2018, Smith2021, Cunha2017})

\begin{itemize}
  \item Signal-to-noise ratio: $\mathrm{S/N} > 100$
  \item Metallicity: $[\mathrm{Fe/H}] > -1$
  \item Effective temperature: $4700\,\mathrm{K} < T_\mathrm{eff} < 5300\,\mathrm{K}$, with uncertainty $\sigma_{T_\mathrm{eff}} < 100\,\mathrm{K}$
  \item Surface gravity: $1 < \log g < 3$, with uncertainty $\sigma_{\log g} < 0.1$
  \item Elemental abundances: all $ -2.2 < [X/\mathrm{H}] < 1$~dex
  \item No flag set `$\mathrm{X\_h\_flags}$' for any element abundance
  \item Elemental abundance uncertainties: all $\sigma_{[X/\mathrm{H}]} < 0.15$~dex
  \item Finite abundance and uncertainty values for all elements
\end{itemize}

\begin{figure}
    \centering
\includegraphics[width=1\linewidth]{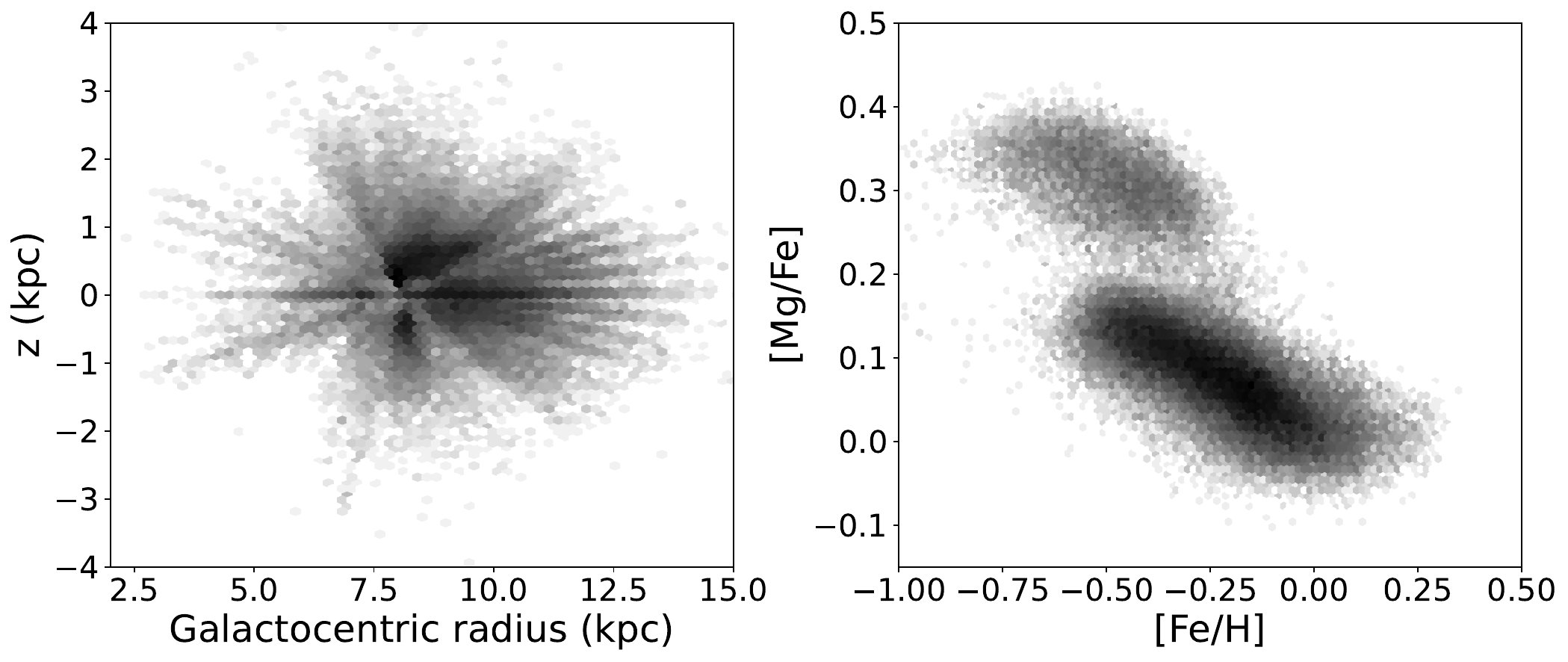}
    \caption{The spatial distribution across \rgal-z and the [Fe/H]-[Mg/Fe] distribution, for the 70,057 selected GG stars.}
    \label{fig:summary}
\end{figure}

This gives a selection of 76,350 red giant stars shown in Figure \ref{fig:summary}. We remove duplicates and a selection of 70,057 remain. We apply parallax and proper motion quality cuts for subsequent analysis using orbital parameters, as described below.

\begin{figure}
    \centering
\includegraphics[width=1\linewidth]{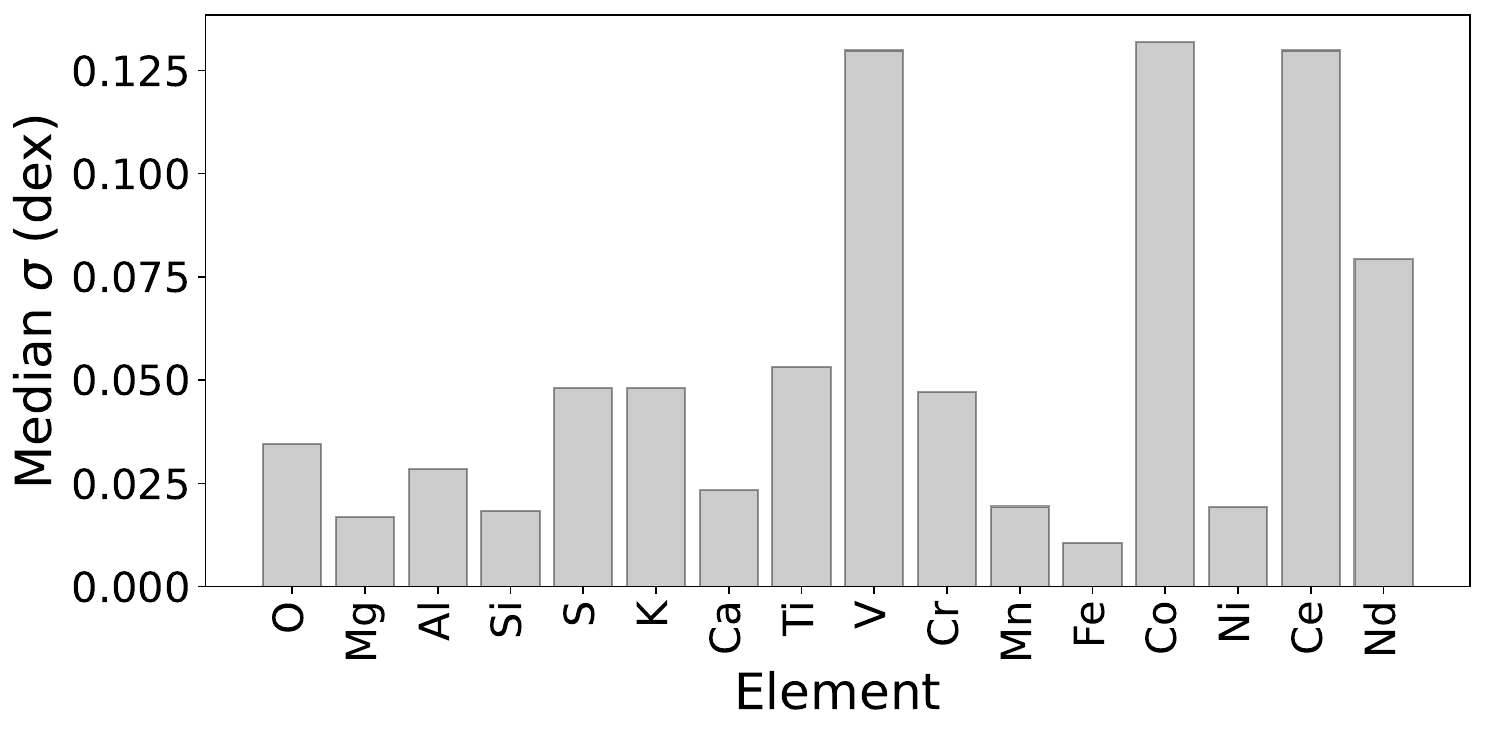}
    \caption{The median uncertainties for all 16 elements in the set of 70,057 Milky Way Mapper stars that meet quality cuts. All elements are with respect to H.}
    \label{fig:errors}
\end{figure}

The stars have 20 measured element abundances [X/H], and we work with a subset of 16. We exclude Na, which previous work has shown can be impacted by a strong diffuse interstellar band \citep{McKinnon2024} and P, which has the highest median uncertainties ($\sigma > 0.15$). We also exclude C and N which change with stellar evolution \citep{Martig2016}. The 16 elements we use, X = Fe,~O,~Mg,~Al,~Si,~S,~K,~Ca,~Ti,~V,~Cr,~Mn,~Fe,~Co,~Ni,~Ce,~Nd, are reported as [X/H]. The median [X/H] uncertainties for the sample are shown in Figure \ref{fig:errors}.

We calculate a number of dynamical parameters: the angular momentum $L_z$, vertical velocity, $v_z$ and the eccentricity, using the \texttt{galpy} package \citep{galpy}, assuming the solar motion from \citet{Schonrich2010}. This is done by combining observed positions and proper motions from \Gaia\ DR3 \citep{GC1}, with radial velocities from Milky Way Mapper, converting them into \texttt{galpy} \texttt{Orbit} objects, and integrating them in the \texttt{MWPotential2014} Milky Way potential model. In addition, the vertical frequency, $\nu_z(R)$, is computed at each star’s Galactocentric radius using the \texttt{galpy} function \texttt{verticalfreq} in the \texttt{MWPotential2014} potential. Our use of a static Milky Way potential follows standard practice for deriving orbital parameters. While this approximation neglects time evolution of the Galactic potential and non-axisymmetric features such as the bar and spiral structure, these effects do not affect the qualitative dynamical trends analysed in this work. The adopted distances are the median geometric distances ($r_{\rm med,geo}$) from \citet{BailerJones2021}.  We use the ages from \citet{StoneM2025} to analyse trends as a function of stellar age.  
In this work, the orbital parameters serve primarily as descriptive diagnostics rather than inputs to a detailed dynamical model.   We exclude $\approx 1000$ stars with distance or proper‐motion uncertainties exceeding 100\%, because these yield unreliable orbital estimates.

\section{Method}
\label{method}

We implement a weighted non-negative matrix factorization to infer latent patterns and weights that generate the abundances. Our aim is to represent each star's element abundance vector as a combination of a small number of characteristic nucleosynthetic patterns, weighted by the star-specific fractional contributions.

\subsection{Generative Formalism}

We represent our abundance matrix as $X$, which has dimensions $N \times K$, where $N$ is the number of stars and $K$ is the number of elements. All entries of $X$ are logarithmic abundances with respect to hydrogen. A shifted abundance matrix $X'$ is constructed from $X$ to enforce non-negativity prior to factorisation, such that $X'_{ij} \ge 0$ for all $i,j$, where $i = 1,\dots,N$ indexes stars and $j = 1,\dots,K$ indexes elements. Each row $X'_i$ contains the  shifted abundance pattern of star $i$.
We then factorise $X'$ as:
\begin{equation}
    X' \approx f\,P ,
\end{equation}
where $f$ is an $N \times M$ matrix of non-negative pattern coefficients and
$P$ is an $M \times K$ matrix of non-negative abundance patterns.
Equivalently, each element of the shifted abundance matrix can be written as
\begin{equation}
    X'_{ij} \approx \sum_{m=1}^{M} f_{im}\,P_{mj}.
\end{equation}

The matrix $f$ ($N \times M$) contains non-negative coefficients describing the contribution of each of the $M$ latent abundance patterns to each star. The coefficients $f$ are dimensionless and, when normalised on a per-star basis, can be interpreted as fractional weights within the model; for interpretability, we normalise the coefficient vectors for each star, and refer to the normalised coefficients as pattern fractions. The matrix $P$ ($M \times K$) contains the corresponding latent abundance patterns (or channels), where each row represents a characteristic combination of elemental abundances. Because the abundance matrix $X$ is expressed in logarithmic units, the patterns encoded in $P$ are defined in log-abundance
space.

 In principle, these latent patterns may correspond to distinct nucleosynthetic signatures arising from specific enrichment processes and star formation histories. However, in practice, the components of $P$ are purely mathematical constructs of the factorisation, and their physical interpretability is not guaranteed. Depending on the input element set, noise structure, and weighting scheme, the decomposition may capture variance-driven or systematic patterns that do not correspond to distinct physical sources. They can also capture multiple but correlated physical processes. As such, while the latent patterns can reflect meaningful enrichment channels, they must be interpreted cautiously and validated against external labels such as age, orbital parameters, or known yield predictions.

We solve for the factorisation using a weighted least-squares loss function that accounts for measurement uncertainties using our shifted abundance matrix to enforce non-negativity, X':

\begin{equation}
\min_{f,\,P \ge 0}
\sum_{i=1}^{N} \sum_{j=1}^{K}
W_{ij}\,\bigl(X'_{ij} - (fP)_{ij}\bigr)^2 .
\end{equation}

Here, $W$ is an $N \times K$ matrix of weights defined as
\begin{equation}
W_{ij} = \frac{1}{\sigma_{ij}^2}
\end{equation}

where $\sigma_{ij}$ denotes the measurement uncertainty of element $j$ for star $i$. We treat the uncertainty covariance matrix as diagonal and neglect inter-element covariances.

\subsection{Implementation}

To satisfy the non-negativity requirement of NMF, we apply a per-star additive shift such that all abundance values are non-negative:

\begin{equation}
{X'}_{ij} = X_{ij} - \min_{j'}(X_{ij'}) + \epsilon
\label{eqn:shift}
\end{equation}

where the minimum is taken over all elements $j'$ measured for star $i$, and $\epsilon$ is a small positive constant added for numerical stability.

This transformation choice is important. Here, we deliberately erase the absolute abundance scale carried by each star, with the goal of instead leveraging the \textit{abundance pattern} (element to element structure). This choice, of shifting each star to a zero-point calibration, therefore removes the absolute abundance scale. We do this because the difference between elements on a per star basis is a physically informative component of the abundances we want to access, as a signature of nucleosynthesis. We verified that using a single global offset (identical for all stars) produces a different set of pattern vectors and markedly reduces spatial and age-dependent contrast in the inferred pattern fractions. Such behaviour indicates that global offsets allow the absolute abundance level to dominate the decomposition. In contrast, the per-star shift emphasises second-order structure after absolute abundance scale which are the differences in abundance pattern morphology across the disc, enabling clearer recovery of relative enrichment behaviour.

We do not standardise (i.e., mean-centre and scale to unit variance) the abundances prior to factorisation because doing so removes the astrophysical variance that anchors the latent patterns to physically meaningful abundance scales. In tests where we standardised each element (subtracting its mean and dividing by its variance), the resulting patterns are mathematically valid, but lose physical interpretability:  NMF components no longer resemble clear nucleosynthetic signatures. Retaining the native abundance variance is therefore important for producing physically grounded pattern vectors and interpretable stellar fraction estimates. This highlights a general property of the matrix factorisation: the recovered components depend on the natural variance, normalisation choices, and the particular set of elements included. Therefore, while the decomposition is stable and meaningful, it is not mathematically unique.

We implement a weighted NMF, updating the
matrices $f$ and $P$ in alternating steps until convergence or until a maximum
of 5{,}000 iterations is reached. Because per-abundance uncertainties must enter
the optimisation explicitly, we use a custom implementation rather than the
standard \texttt{sklearn} NMF routine. Our implementation takes as input the
abundance matrix $X$, the associated weight matrix $W$, and the user-specified
number of components $M$.

The factors are obtained by minimising the weighted squared reconstruction loss
under non-negativity constraints:
\begin{equation}
\mathcal{L}(f, P) =
\sum_{i=1}^{N} \sum_{j=1}^{K}
W_{ij}\,\left( X'_{ij} - \sum_{m=1}^{M} f_{im} P_{mj} \right)^2 .
\end{equation}

We optimise this loss using the standard multiplicative update rules of
\citet{Lee2001}, which preserve non-negativity and guarantee a monotonic decrease
in the reconstruction error. These updates alternately optimise $f$ and $P$. The update equations are:

\begin{equation}
P \leftarrow P \cdot
\frac{(W \cdot X')^\top f}{(W \cdot (fP))^\top f + \epsilon}
\quad \text{and} \quad
f \leftarrow f \cdot
\frac{(W \cdot X') P^\top}{(W \cdot (fP)) P^\top + \epsilon}
\end{equation}
where $\cdot$ denotes element-wise multiplication and $\epsilon$ is a small
constant added for numerical stability.

The matrix $W$ is used to weight each squared residual by its measurement uncertainty.

Physically, this model assumes that each observed abundance vector arises from a sum of non-negative contributions from a small number of latent nucleosynthetic patterns, expressed in $P$, with star-specific amplitudes given by $f$. The product $fP$ then reconstructs the abundance pattern per star. The success of the model can be interpreted as both a data-driven identification of dominant patterns, generated by one or more sources, and a test of whether a low-dimensional representation can explain observed chemical diversity. Applying this method is motivated by prior work that shows that a low-dimensional representation does explain a substantial fraction of the chemical diversity \citep{Casey2019}. Previous work has demonstrated that stellar abundances occupy a low-dimensional space and are highly predictable from a small subset of elements \citep[e.g.][]{Ness2022, Mead2025}, while physically motivated models have inferred source fraction coefficients under assumed nucleosynthetic constraints \citep[e.g.][]{Weinberg2019, Griffith2021, Griffith2024}. In contrast, our approach learns a small number of shared, additive enrichment patterns directly from the data,
without imposing any assumptions about the underlying generating sources.

\subsection{Model Evaluation and Selection of Number of Latent Patterns}
\label{results1}

As the number of latent variables $M$ increases, the generative model improves its ability to reconstruct observed abundances. This is expected; more degrees of freedom allow for better approximation of the data. Once  $M$  approaches the number of elements, $K$, the model becomes degenerate, as each element can be trivially represented by its own latent channel.  In general, this is a tradeoff; a smaller number of components is easier for interpretation, but a larger number better generates the abundances. To avoid overfitting and preserve interpretability, we seek the smallest number of latent components that yield satisfactory reconstruction quality. 

We evaluate the model performance under different values of $M$ using the distribution of the per-star $\chi^2$, which quantifies how well the generative model reproduces the observed abundances of each individual star. Although the non-negative matrix factorisation is performed on abundances that are shifted to enforce positivity, all reconstructions are transformed back to the original measurement scale by re-applying the per-star abundance offset prior to evaluating the goodness of fit. The per-star $\chi^2$ is then computed using the same form as the weighted loss function that was minimised to solve for both the latent abundance pattern matrix $P$ and the per-star coefficient matrix $f$:

\begin{equation}
\chi^2_i =
\sum_{j=1}^{K} W_{ij}
\left[
X_{ij}
-
\left(
(fP)_{ij}
+ \min_{j'}(X_{ij'})
- \epsilon
\right)
\right]^2 .
\label{eqn:chisquared}
\end{equation}

In practice, we find that four to six latent components offer a useful trade-off between explanatory power and simplicity. This balance permits physical interpretation of each latent channel in terms of nucleosynthetic sources, while maintaining sensitivity to residuals that may reflect evolutionary diversity or observational uncertainties.

\begin{figure}
    \centering
    \includegraphics[width=1\linewidth]{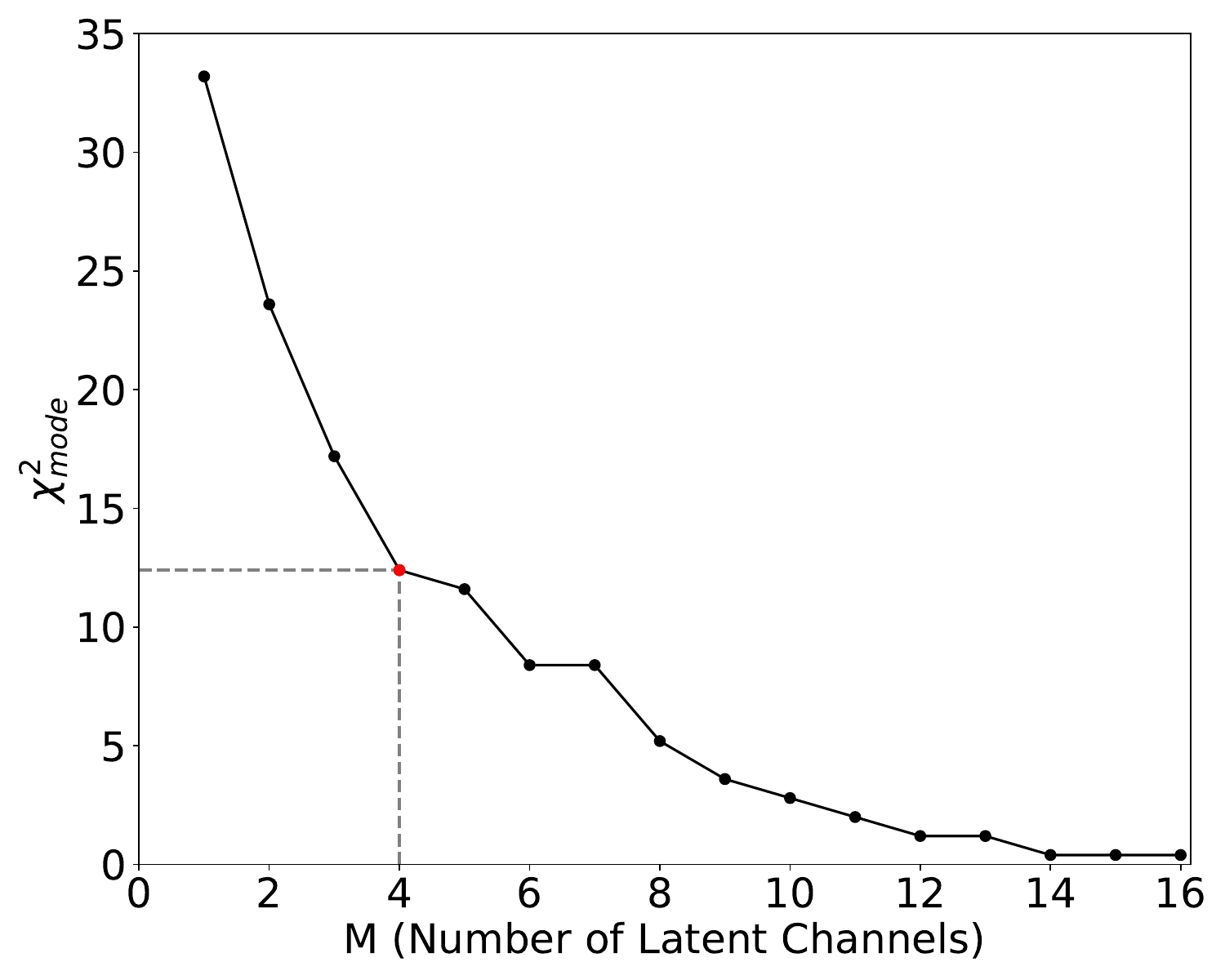}
\caption{The mode of the distribution of per-star \(\chi^2\) between the generated and \textsc{ASTRA ASPCAP} abundances for the 70,057 Milky Way Mapper disc stars, shown as a function of the number of latent components (pattern vectors) used in the model of the 16 element abundances. We adopt $M=4$ components for our analysis.}
 \label{fig:mode}
\end{figure}

Figure~\ref{fig:mode} shows the mode of the per-star $\chi^2$ distribution, highlighting the most common reconstruction quality across the sample. Unlike the population-wide reconstruction error, which can be skewed by a minority of poorly fitted stars, the mode better reflects typical performance for the majority.  For $M=4$ latent components, the mode of the reduced $\chi^2$ (the $\chi^2$ divided by the degrees of freedom, i.e.\ 16 elements minus 4 patterns) is $\sim 1$. At $M=5$, the reduced $\chi^2$ is also essentially unity, the textbook definition of a statistically optimal fit. However, the improvement relative to $M=4$ is marginal. We therefore adopt $M=4$ as our fiducial basis: it balances statistical quality with interpretability, allowing finer separation of nucleosynthetic groups (particularly those associated with massive stars) while keeping the number of patterns tractable.  
Our main results and conclusions remain qualitatively unchanged for $M=3$ or $M=5$. In practice, the ``best fit'' depends not only on statistical criteria but also on which elements are available and their relative measurement precision.

\begin{figure}
    \centering
\includegraphics[width=1\linewidth]{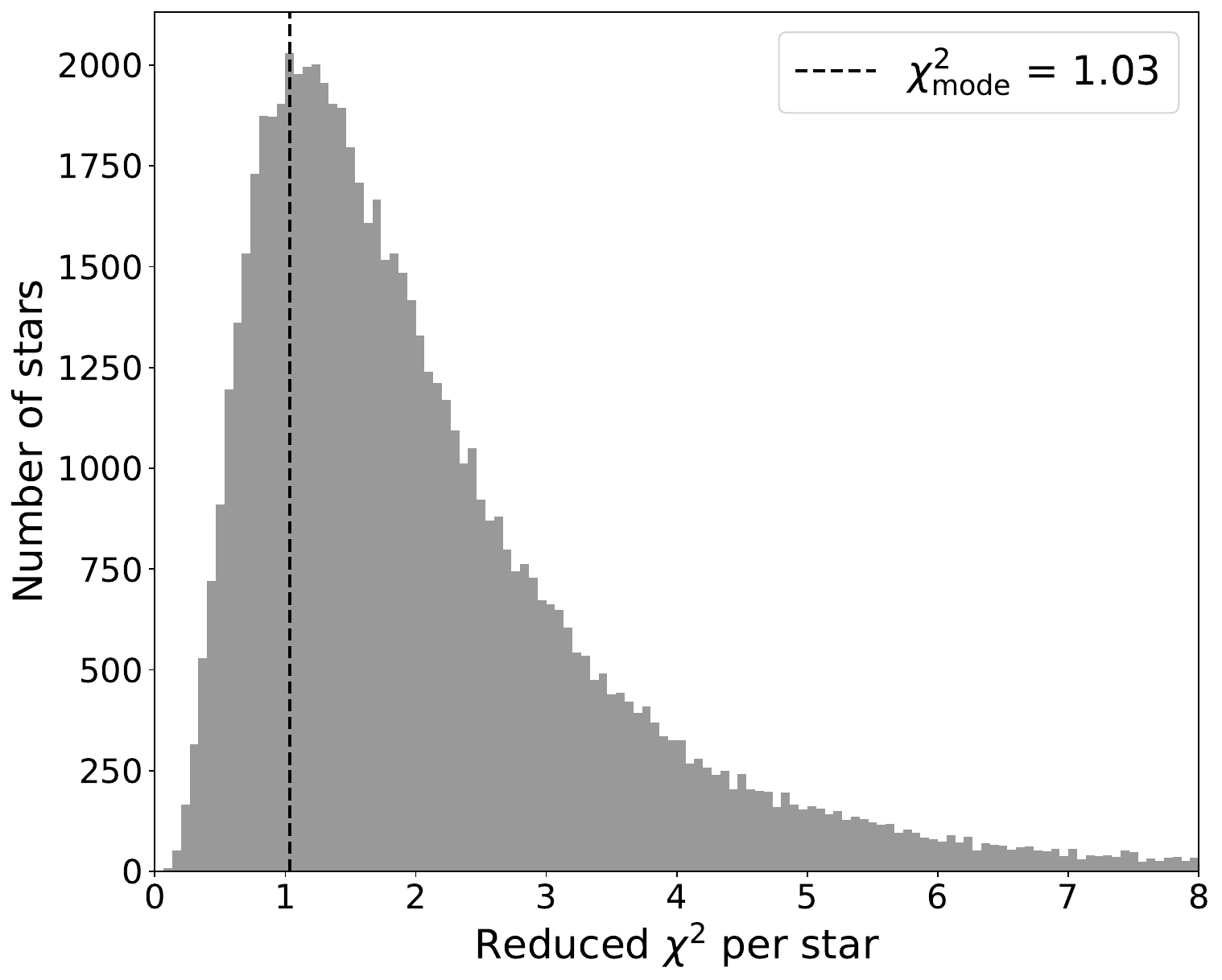}
\caption{Distribution of per-star reduced \(\chi^2\) values for the sample of 70,057 stars, based on a latent variable model with  $M=4$ components (patterns).}
\label{fig:chi2}
\end{figure}

In Figure \ref{fig:chi2}, we show the reduced $\chi^2$ distribution for the $M=4$ latent vector model. The distribution peaks at a mode of $\chi^2_{\text{red}} \sim 1$, corresponding to a total $\chi^2 \sim 12.4$ with 12 degrees of freedom. Approximately 20\% of stars have a reduced $\chi^2$ $>$ 3 and $\sim$5\% of stars have reduced $\chi^2$ $>$ 5.

\begin{figure*}
    \centering
\includegraphics[width=1\linewidth]{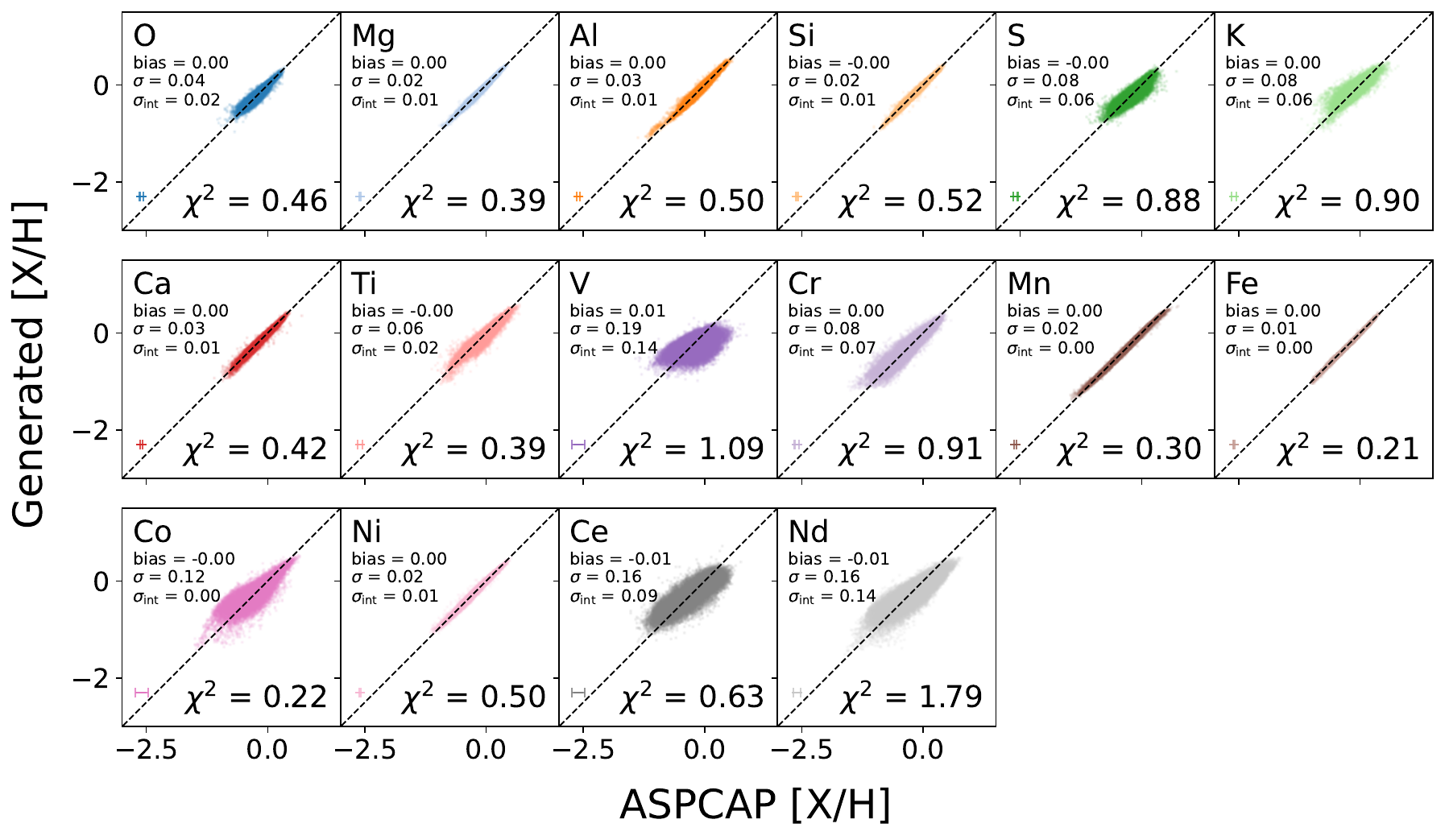}
    \caption{Elemental abundances predicted by the latent variable model for the Milky Way Mapper disc sample, using $M = 4$ latent components. The measured \textsc{ASTRA ASPCAP} abundances are on the x-axis and the generated abundance using the NMF basis are shown on the y-axis. The bias, scatter and intrinsic scatter (subtracting the mean error in quadrature) are indicated in each sub-figure. The abundances are reconstructed from the solved pattern matrix $P$ and coefficient vectors $f$ via $X=f P$, with the reconstruction corrected for the per-star abundance offset.}
 \label{fig:elemfits}
\end{figure*}

In Figure~\ref{fig:elemfits}, we show the predicted elemental abundances reconstructed using the solved pattern matrix $P$ and coefficient vectors $f$; the bias, scatter and intrisic scatter (whereby the mean uncertainty of each element is subtracted in quadrature form the scatter) are shown in each panel. In general, the fits are precise for the $M=4$ latent basis, although the elements with the highest observational uncertainties -- V, Ce, and Nd -- are also those with the largest residuals and least well-fit. This reflects that their astrophysical production channels are not as cleanly separable in the latent basis given higher uncertainties. Although the reconstruction for these elements is noisier, the bulk of the elements are well fit, providing confidence in the robustness of the overall decomposition.

\begin{figure*}
    \centering
\includegraphics[width=1\linewidth]{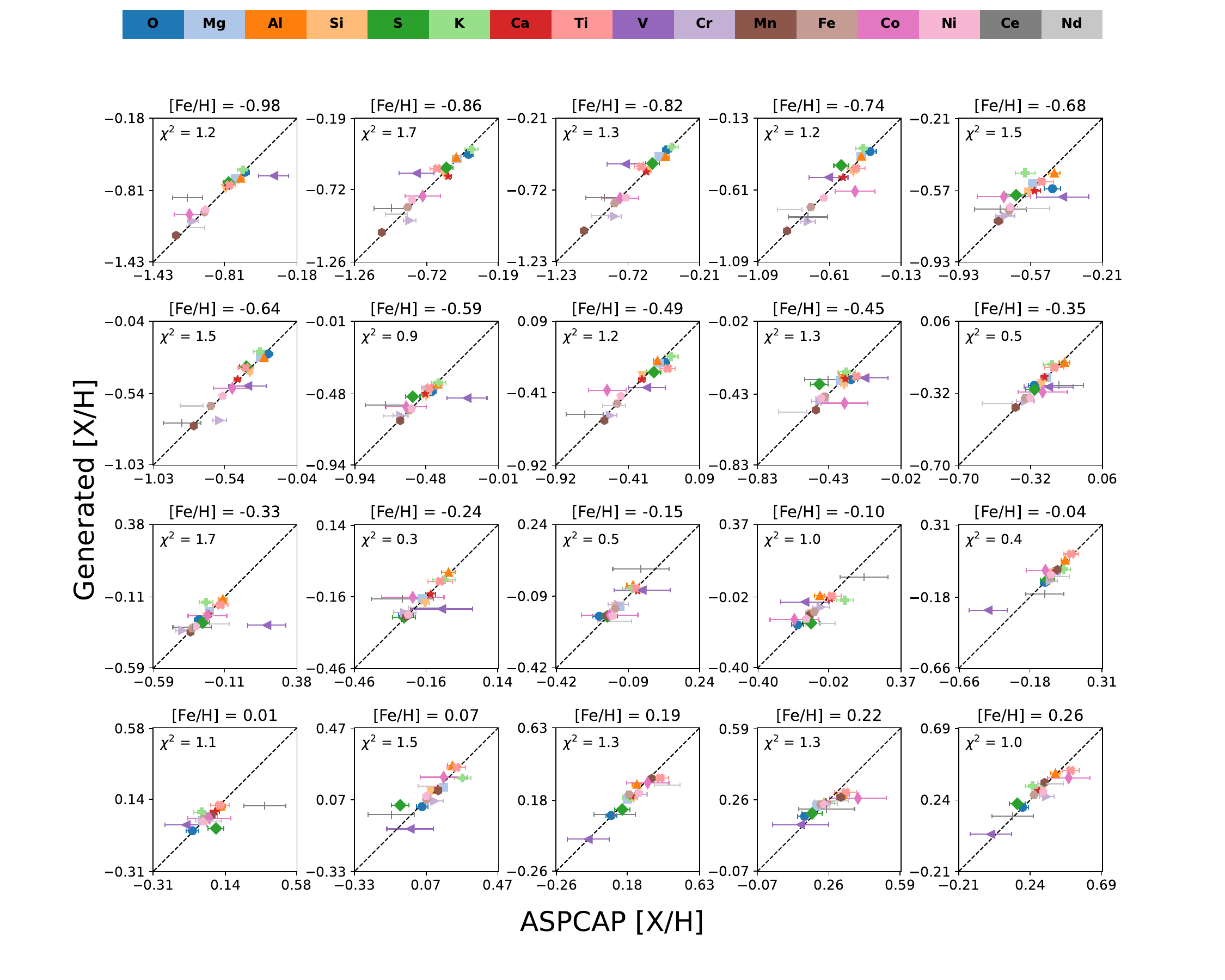}
    \caption{An example of the reported DR19 \textsc{ASTRA ASPCAP} abundances for individual stars, versus the abundances generated with the latent vector model. Each panel shows a star that is randomly selected from the set of $\sim$40,660 stars with reduced $\chi^2$ $<$ 2, from within narrow metallicity ranges in each sub-panel, to represent the full range of the sample across all panels. The stars are ordered from metal-poor to metal-rich.  Approximately 58\% of the sample have this quality of fit with a reduced $\chi^2$ $<$ 2.}
    \label{fig:starfits}
\end{figure*}

In Figure~\ref{fig:starfits}, we show example model-generated abundances compared to the measured \textsc{ASTRA ASPCAP} abundances for a random subset of stars, ordered by [Fe/H] and selected to have reduced $\chi^2 < 2$; this quality cut is met by approximately 58\% of the sample (with $\approx80$\% meeting a reduced $\chi^2 <$ 3). For each element, we calculate and display the bias, total scatter, and estimated intrinsic scatter relative to the model predictions.

\section{Results}

\subsection{Latent Patterns as Enrichment Channels}
\label{validation}

After the NMF decomposition, the matrix $f$ contains the coefficients describing the contribution of each latent component to the composition of each star, while the matrix $P$ encodes the corresponding abundance patterns. Each component is characterised by a distinct set of per-element amplitudes in $P$, which quantify the relative contribution of individual elements. We associate these patterns with contributions from different physical sources, and subsequently refer to these latent components as \textit{channels}. Each channel is defined by a characteristic abundance pattern, associated with an underlying nucleosynthetic source.

\begin{figure*}
    \centering
\includegraphics[width=1\linewidth]{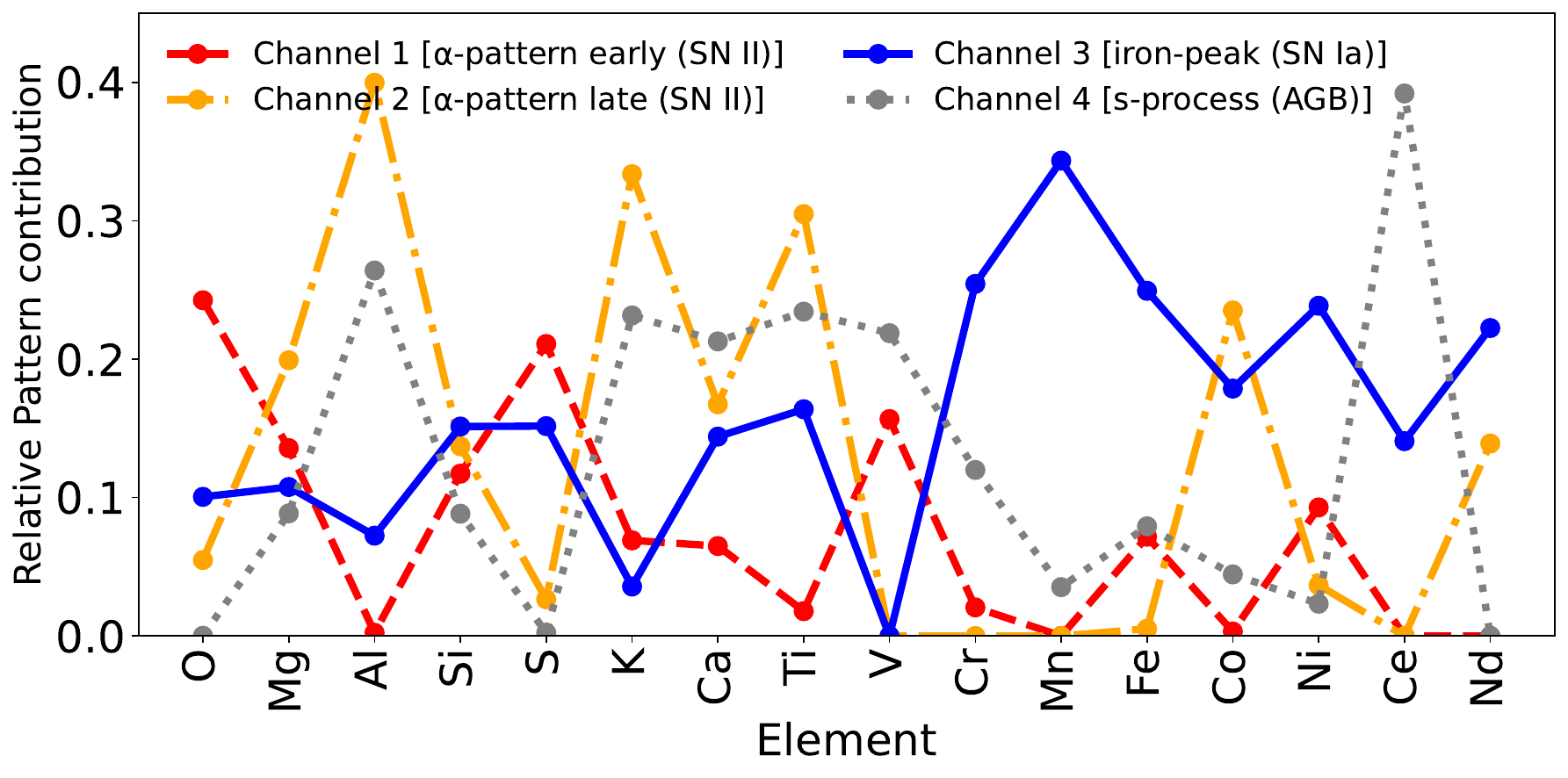}
\caption{The four latent pattern vectors, $P$, identified from data, showing the relative contribution of each element to each pattern. These patterns do not represent single yields, but rather integrated enrichment signatures. They each show distinct behaviour and can subsequently be associated with dominant enrichment channels or sources, as indicated in the legend. Their physical interpretation is guided by canonical tracer elements, but this does not imply unique production of all contributing elements by a single source. Channel 1: Early SN II pattern–enrichment from massive core-collapse supernovae producing O, Mg, and S. Channel 2: Late SN II pattern–lower-mass core-collapse events producing Al, Si, K, Ca, Ti. Channel 3: Iron-peak pattern–thermonuclear SN Ia contributing Cr, Mn, Fe, Co, Ni on longer timescales. Channel 4: Delayed-enrichment/s-process–dominated pattern with a Ce peak, from primarily AGB stars returning heavy elements such as Ce through winds, together with correlated structure in Ca, Ti, V, and K, likely reflecting metallicity-dependent yields rather than direct AGB production. The physical validity of each channel is tested using independent indicators (e.g. age, position, orbits).}
 \label{fig:patterns}
\end{figure*}

Figure~\ref{fig:patterns} displays the $M = 4$ latent patterns, $P$. The pattern values shown in Figure \ref{fig:patterns} have not been normalised; they reflect the relative contribution of each element to each channel, as derived from the model. We interpret each pattern as being associated with a particular nucleosynthetic channel. We emphasise that these patterns are \textit{not} direct yield vectors. Rather, they reflect the mixed contributions of multiple nucleosynthetic sources over varying timescales of star formation. Each channel subsequently represents a composite enrichment pattern, which can be associated with a dominant source based on the elements that contribute most significantly. At the same time, we note that latent patterns can also capture non-physical structures, for example covariance introduced by measurement uncertainties, survey systematics, or correlated errors among particular elements. Caution is therefore required: while these empirical patterns provide a powerful tool to separate enrichment processes operating on different timescales, they should not be interpreted as one-to-one maps of pure source contribution, or nucleosynthetic yields.

\begin{itemize}

\item Channel~1 represents an ``early'' $\alpha$-pattern, dominated by strong contributions from O, S and Mg, which are elements produced in hydrostatic fusion in massive stars \citep{Woosley1995, Blancato2019}. Their yields are highly sensitive to progenitor mass, making this channel a signature of enrichment from the most massive, short-lived SN~II \citep{Nomoto2013}. At the same time, this channel also shows  contributions from iron-peak elements including Fe and Ni, which must co-vary with O, S and Mg in the data but are not consistent with the same hydrostatic, high-mass nucleosynthetic origin. This highlights that the latent patterns are not pure source or yield vectors, but empirical co-variance structures that can mix physical signals with other correlated trends.

\item Channel~2 is a ``late'' $\alpha$-pattern, with strong contributions from Al, K, Ti, Ca, Mg, and Si. Elements including Si, Ca, and Ti are produced predominantly during explosive O and Si fusion in SN~II, with yields that peak in lower-mass massive stars \citep{Woosley1995, Nomoto2006, CK2020}.  This suggests the channel  is consistent with a later phase of core-collapse enrichment relative to channel~1.  Similarly to channel~1, the presence of elements like  Al, Co and Nd likely reflects covariance in the data rather than a pure nucleosynthetic source.

\item Channel~3 shows a large contribution from the iron-peak elements including Cr, Mn, Fe, Co, and Ni. These elements are characteristic of enrichment by Type~Ia supernovae (SN~Ia) \citep{Nomoto1997, Kobayashi2020}, consistent with the interpretation of this channel as the  delayed iron-peak contribution from low-mass stellar explosions. The $r$-process element Nd also projects onto this channel, not necessarily because it shares the same nucleosynthetic origin, but likely because of a similar enrichment timescale from longer-lived, lower-mass progenitors. The s-process element Ce similarly is a strong contribution in this channel along with Al and Ti. This again indicates that elements that express together can always reflect covariance rather than direct physical association.

\item Channel~4 shows a Ce peak, as well as large contributions from V and Al. The  Ce feature is a clear signature of the slow neutron-capture  process ($s$-process) contributions from Asymptotic giant branch (AGB) stars The associated odd-$Z$ elements may reflect weaker AGB contributions or (non-physical) empirical covariance e.g., from measurement uncertainties or survey systematics. We therefore interpret this channel as tracing enrichment from low- and intermediate-mass AGB stars, while cautioning that not all contributing elements in this channel necessarily share the same AGB origin.

\end{itemize}

Although these are not `pure' representations, in the subsequent analysis, we therefore associate channel~1 with early $\alpha$-element enrichment from high-mass SN~II\footnote{We use the term SN~II as a compact label for the dominant enrichment from the core collapse of massive stars. Core collapse supernovae events also include Type Ib and Ic and therefore this term is strictly a convenient shorthand to be consistent with the other channel label nomenclature.}, channel~2  with late $\alpha$-element enrichment from the broader SN~II mass range, channel 3 with iron-peak enrichment from SN~Ia, and channel~4 with $s$-process enrichment from AGB stars.  While the channels are not pure tracers of individual nucleosynthetic processes, the fact that they separate into coherent patterns allows us to meaningfully connect them to physical enrichment sources. These groupings reflect the dominant behavior of elements across the stellar population, shaped by overlapping contributions from multiple processes acting over Galactic timescales. We subsequently test for and validate the physical plausability of this association, but emphasise the latent decomposition remains a powerful means of leveraging joint abundances even without the nucleosynthetic pathway interpretation.

We highlight that the particular pattern decomposition seen here is not a fundamental and unique property of the Galaxy, but rather a property of the input data and modelling choices. The detailed form of the patterns will vary with changes in the adopted abundances and their uncertanities, normalisation, or weighting scheme in the NMF. Nevertheless, the method is a useful tool for separating processes that operate on different enrichment timescales. We have checked alternative dimensionalities (e.g.\ $M=3$ and $M=5$), different element selections, and different uncertainty-weighting formalisms, all of which yield qualitatively similar results (with the exception of excluding key elements like Ce, where some enrichment granularity is subsequently lost). Therefore, while the precise decomposition depends on the details of the analysis, the broad  conclusions reported here are robust. This approach also effectively de-noises the abundances; the reduced inverse-variance weighted uncertainties of the combined elements into patterns is $\leq 0.01$ each.

\subsection{Stars not generated by the model}
\label{failures}

\begin{figure}
    \centering
       \includegraphics[width=1\linewidth]{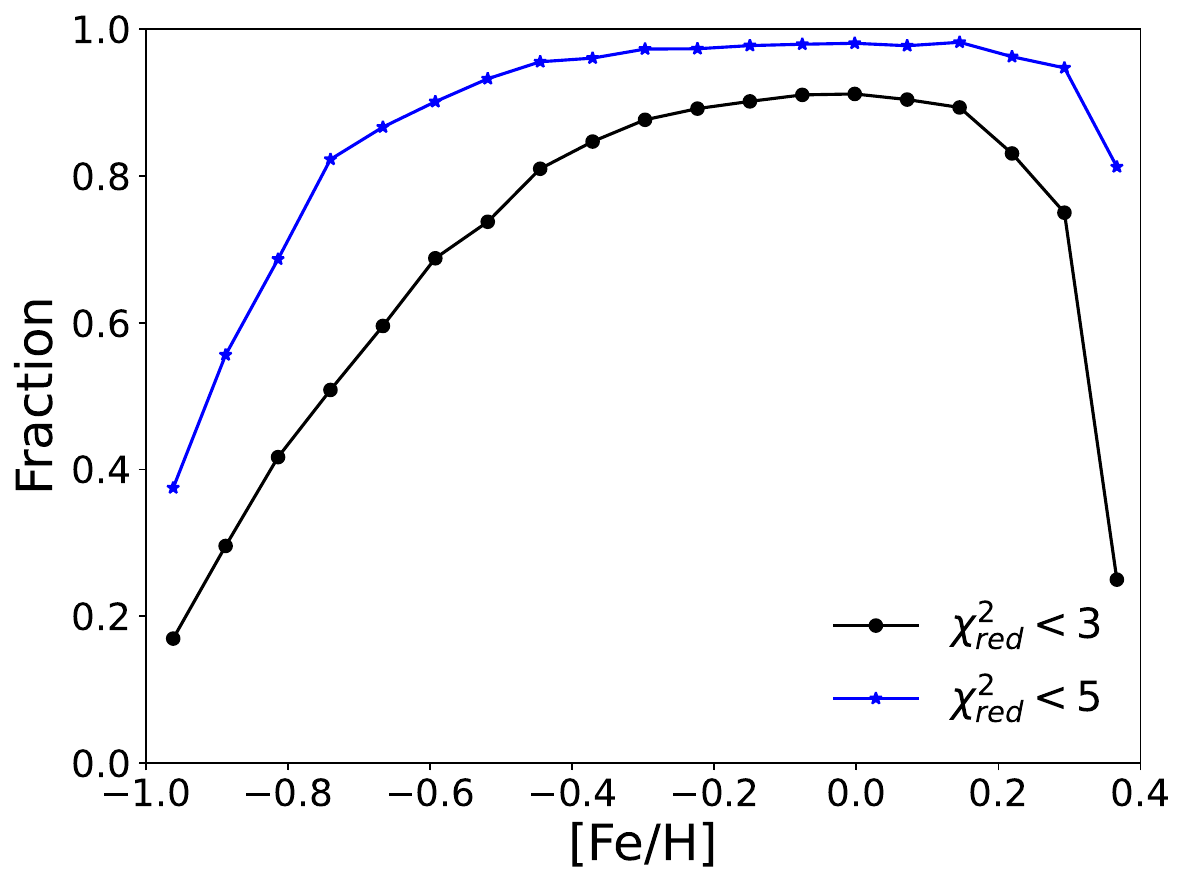}
    \caption{The distribution of the goodness of the generated model fits compared to the data, parameterised using a reduced $\chi^2$ metric. The y-axis reports the fraction of stars as a function of metallicity with a reduced $\chi^2$ $<$ 3 and reduced $\chi^2$ $<$ 5, respectively. The edges of the distribution are poorly fit by the model. Increasing the number of latent channels, $M$ broadens the distribution slightly, but the overall result is consistent with what is seen for $M=4$ even when including additional latent variables. Substantial model deterioration happens around [Fe/H] $< -0.6$ and $> 0.2$ and only 10\% of the stars show reasonable fits at [Fe/H] $<$ -0.8 and [Fe/H] $>$ 0.3.}
    \label{fig:feh_fraction}
\end{figure}

As shown in Figure~\ref{fig:elemfits}, the model can reproduce the stellar
abundances of the majority of the disc population, with 95\% of stars having a
reduced $\chi^2 < 5$. To generate each star’s abundance vector, we multiply the pattern matrix $P$ by the coefficient matrix $f$ and add the offset introduced to enforce non-negativity. We then calculate the $\chi^2$ of the true versus generated abundances (Equation \ref{eqn:chisquared}). Notably, there are stars that are well-fit across the entire metallicity range of the disc, from $-1 <$ [Fe/H] $< 0.3$. However, the overall performance of the model rapidly declines at [Fe/H] $< -0.6$, and [Fe/H] $> 0.2$.  This is shown in Figure \ref{fig:feh_fraction}. While at $-0.6 <$ [Fe / H] $< 0.2$ $\sim$ 80\% stars have a reduced $\chi^2 < 3$, only 40\% of stars have a reduced $\chi^2 < 3$ at [Fe/H] $> 0.3$, and only 25\% have $\chi^2 < 3$ at [Fe/H] $<$ -0.8.

This decline in performance at the metallicity extremes may reflect both model limitations, in particular the restricted flexibility of a linear formalism, and genuine astrophysical differences. While a linear model cannot reproduce non-linear abundance pattern behaviour, one way such a limitations would manifest is in produce poor fits across the sample. Another modelling limitation is the relative sparsity of stars at very low and very high [Fe/H]. This means that the NMF basis is primarily constrained by the bulk of the disc population, and therefore less effective at capturing the tails of the distribution. However, failure of the model in these regimes remains informative, as it does not necessarily indicate an intrinsic limitation of the method, but rather that these stars may represent chemical enrichment processes that are less well captured by the dominant variance modes of the disc. We therefore examine poorly fit stars, to understand if they also show distinct kinematics compared to the rest of the disc, providing an independent way of testing if the deviations might reflect different populations, or not.

\begin{figure}
    \centering
      \includegraphics[width=1\linewidth]{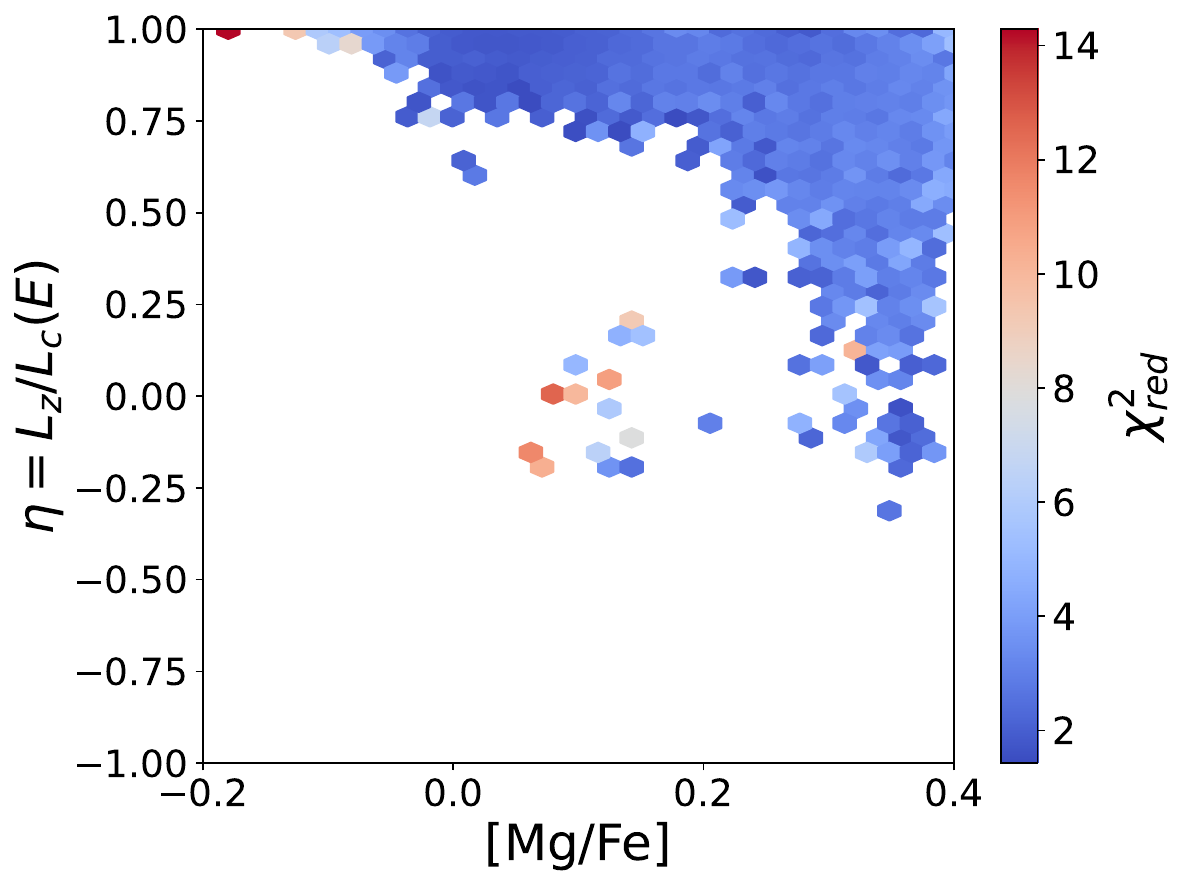}
    \caption{A two-dimensional hex-binned diagram of stars in orbital circularity ($\eta = L_z / L_c(E)$) versus $\mathrm{[Mg/Fe]}$, coloured by the mean reduced chi-squared ($\chi^2_{\rm red}$) from the latent-abundance model (only bins with $\geq 3$ stars are shown). Stars with high $\chi^2_{\rm red}$ are broadly distributed, but their average values are systematically higher in specific regions of parameter space. In particular, stars near $\eta \approx 0$ show elevated $\chi^2_{\rm red}$, consistent with an accreted component at relatively low $|z|$ ($\lesssim 5$ kpc). }
 \label{fig:etamap_chi2}
\end{figure}

In total, we find that $\approx$5\% of stars have a reduced $\chi^2$ greater than 5, which we consider to be not well generated by the model and which we exclude from subsequent analysis. 

Stars with a reduced $\chi^2 > 5$ are found across the phase space of the sample, with concentrations in particular regions.  We investigate the distribution of the goodness of fit in chemo-dynamical space by first examining the plane of orbital circularity, $\eta \equiv L_z / L_c(E)$, versus [Mg/Fe]. The orbital circularity $\eta$ is calculated using \texttt{galpy} by integrating each star's orbit in the \texttt{MWPotential2014} Galactic potential and adopting the median geometric distance `$r\_{med}\_{geo}$' \citep{BailerJones2021}. The angular momentum in the $z$ direction, $L_z$, is compared to the angular momentum of a circular orbit at the same total energy, $L_c(E)$. This yields a diagnostic of orbital type: $\eta \approx 1$ for stars on circular disc-like orbits, $\eta \sim 0$ for eccentric orbits, and $\eta < 0$ for retrograde orbits. The $\eta$- metallicity plane is analysed in \citet{Chandra2023} in their three-phase disc analysis.

Figure~\ref{fig:etamap_chi2} shows a binned histogram of $\sim$65,313 stars, selecting the $\sim$95\% of stars with distance uncertainties $<30$\% and heights from the plane $|z| < 5$~kpc from the full parent sample of 70,057 stars. These are shown binned and coloured by their reduced $\chi^2$ value, $\chi^2_{\rm red}$, from the latent model fit to their  abundances. The overall distribution of the disc in the $\eta$–plane is consistent with that reported by \citet{Chandra2023}, with stars on more circular orbits dominating at lower [Mg/Fe], corresponding on average to younger, more metal-rich populations. The regions of elevated mean $\chi^2_{\rm red}$, highlighted in red, indicate subsets of stars that are poorly described by the model. One such feature appears around $(\mathrm{[Mg/Fe]}, \eta) \sim (0.1, 0)$, which likely traces the accreted material in the Milky Way disc, for example, the remnants of disrupted dwarf galaxies Gaia-Enceladeus-Sausage \citep[e.g.][]{Belokurovsplash, Helmi2020, Naidu2020}.

This map also shows that in general, the reduced $\chi^2$ shows a good population fit across the entire disc plane, from more radial to circular orbits, with exception the most [Mg/Fe] poor tail of the distribution.

\begin{figure}
    \centering
\includegraphics[width=0.9\linewidth]{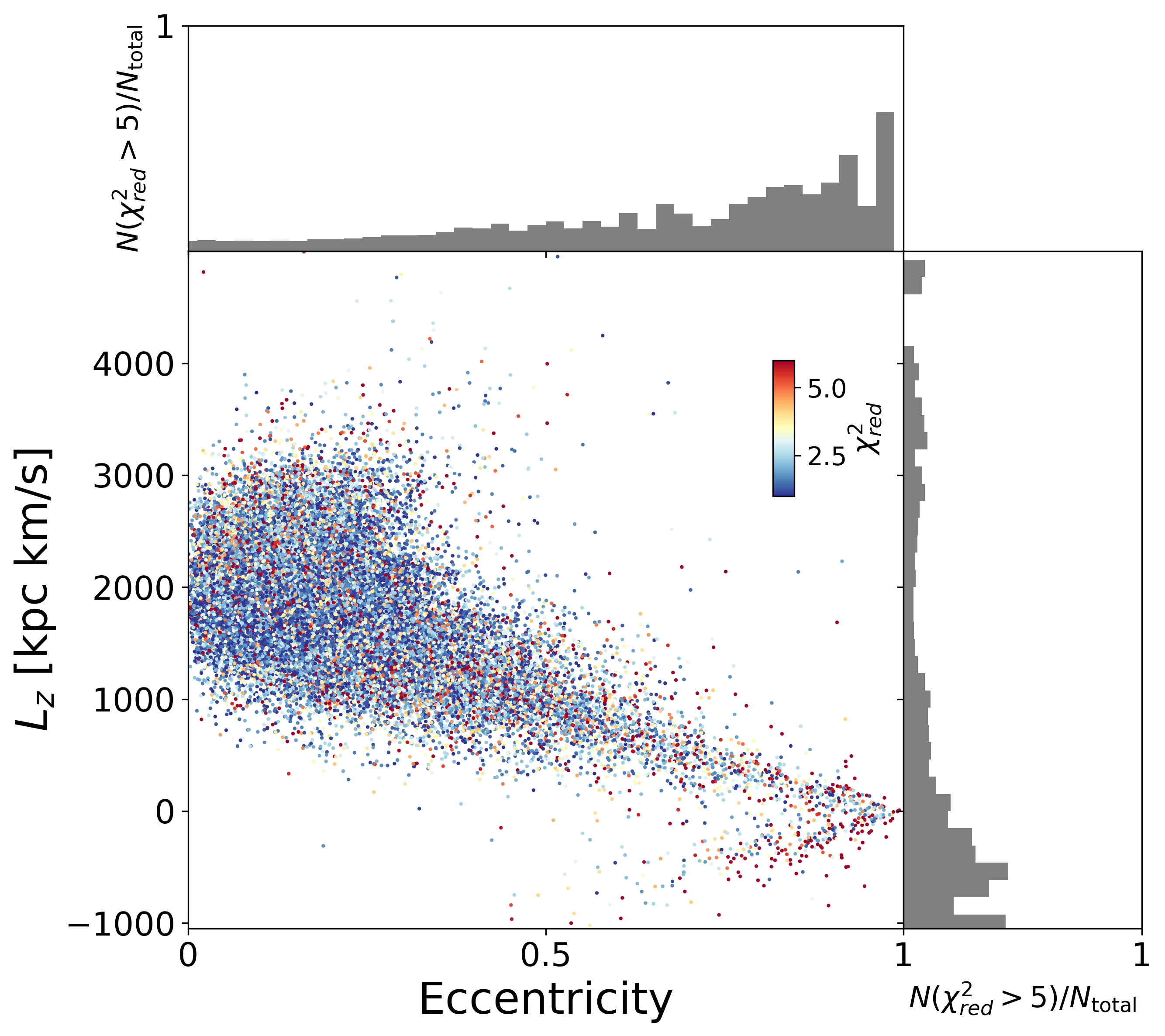}
\includegraphics[width=0.9\linewidth]{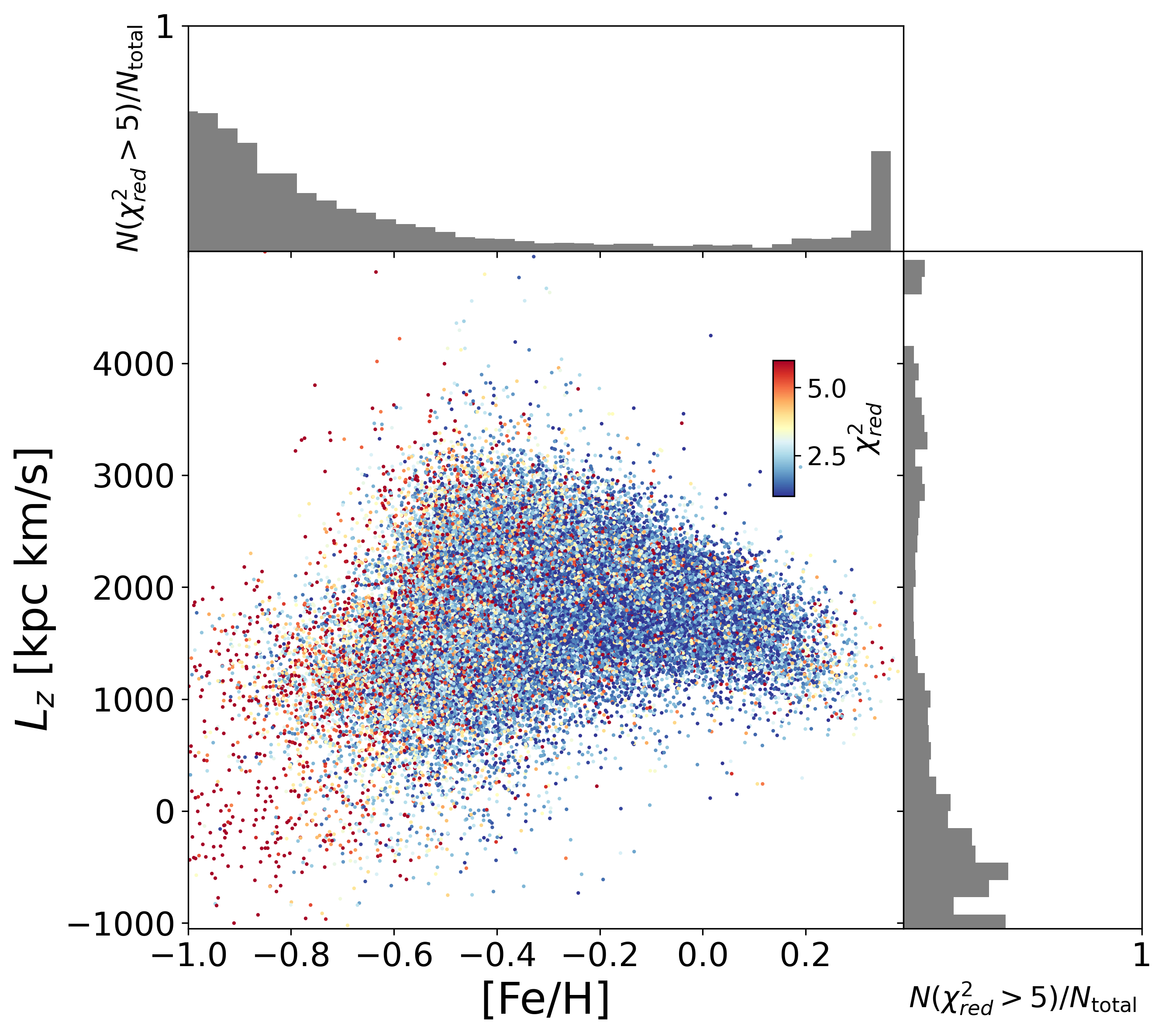}
    \caption{Identifying the parameter space of the stars with abundances that are poorly generated by the latent pattern model. Stars with poor fits, which are defined as (reduced) $\chi^2_{\mathrm{red}}>5$, are  reported in the histograms, which show where this small $\sim$5\% subset lies in parameter space by measuring the fraction of (reduced) $\chi^2_{\mathrm{red}}>5$ fits per bin. The failure rate is highest ($\approx$30\%) at the metallicity extremes of $[\mathrm{Fe}/\mathrm{H}]<-0.6$ and $[\mathrm{Fe}/\mathrm{H}]>+0.3$. Dynamically hot, low--angular-momentum orbits are also over-represented: for stars with eccentricity $> 0.5$ and $L_z<1000$ (including retrograde, $L_z\le 0$), $\approx 45\%$ exceed the $\chi^2_{\mathrm{red}}>5$ threshold.}
    \label{fig:outliers}
\end{figure}

We show additional projections that demonstrate the dynamical coherence of stars that are poorly fit by the model seen in Figure \ref{fig:etamap_chi2}, in Figure~\ref{fig:outliers}. The figure reports the fraction of stars with $\chi^2_{\rm red} > 5$ across angular momentum–eccentricity space (and in the central scatter plot these are the red points). This shows the regions in this plane where poor fits are most prevalent, also seen as groups of stars concentrated in Figure \ref{fig:etamap_chi2} around $(\mathrm{[Mg/Fe]}, \eta) \sim (0.1, 0)$ which corresponds to the metal-poor stars with poor fits $\chi^2 >$ 5 in Figure \ref{fig:outliers}. Figure~\ref{fig:outliers} more clearly identifies the different regions of phase space where stars with the highest fraction of poor abundance predictions are found. These include the most metal-rich stars in the sample ([Fe/H] $>0.3$) and the more metal-poor stars ([Fe/H] $<-0.6$); among the latter, many also have low angular momentum ($|L_z|<1000$) and high eccentricities. The histograms along each axis show the fraction of stars that are poorly fit by the model. Of the $\sim$5\% of stars with $L_z<1000$, 15\% (40\%) are poorly fit, with $\chi^2_{\rm red}>5$ (3). Again, some contribution from edge effects or sparsely sampled regions of the abundance space is expected, particularly at the metallicity extremes. However, the dynamical coherence of these stars shows in many cases, that the failures reflect populations that may not share the dominant enrichment pathway of the disc.

In combination with their dynamics, this suggests an accreted origin for a subset of the model failures, although stars in this parameter space have also been linked to an in-situ “Splash” population disturbed by early mergers \citep{Belokurovsplash, Orti2023}. Abundance evidence supports the interpretation that some of these are accreted stars. About 10\% of poorly fit stars have [Al/Fe] $<0$, consistent with previously identified accreted populations \citep{Belokurov2022, Feuillet2022}, while another $\sim$10\% have [Mg/Fe] $<0$, again indicative of dwarf-galaxy origins. The majority of poorly fit stars lie at the metallicity extremes of the disc ([Fe/H] $>0.3$ or $<-0.6$) (see the bottom panel of Figure~\ref{fig:outliers}).

The poor fits are therefore informative: they identify populations whose enrichment histories diverge from the dominant Galactic pathways captured by the latent basis. In particular, the sharp decline in model success at low metallicity, alongside the dynamical confinement of a fraction of these stars, argues against the metal-poor extent of the sample being a chemically uniform extension of the metal-rich disc.  That is, while the model can can generate the abundances of disc stars down to the lowest metallicty [Fe/H] $\sim -1$, poor fits are in much larger fraction below [Fe/H] $< -0.6$. These appear across the dynamical extent of the disc, but show highest concentration at high eccentricty and low angular momentum (i.e. see Figure~\ref{fig:feh_fraction} and Figure~\ref{fig:outliers}). Therefore, the data are more naturally explained by a mixture of distinct origins below [Fe/H] $< -0.6$. This includes accreted material, clumpy early star formation, or spatially inhomogeneous gas inflow. We therefore interpret this behaviour as reflecting genuine diversity in star formation and enrichment channels rather than limitations of the latent factor model alone. This interpretation is supported by the persistence of the trend as model dimensionality increases: while the overall reconstruction improves, a subset of stars remains systematically poorly fit. Smooth metallicity-dependent yield variations alone are therefore unlikely to account for this behaviour, as increasing the number of latent components would be expected to absorb such effects rather than leave a stable off-manifold population.

\begin{figure}
    \centering
        \includegraphics[width=1\linewidth]{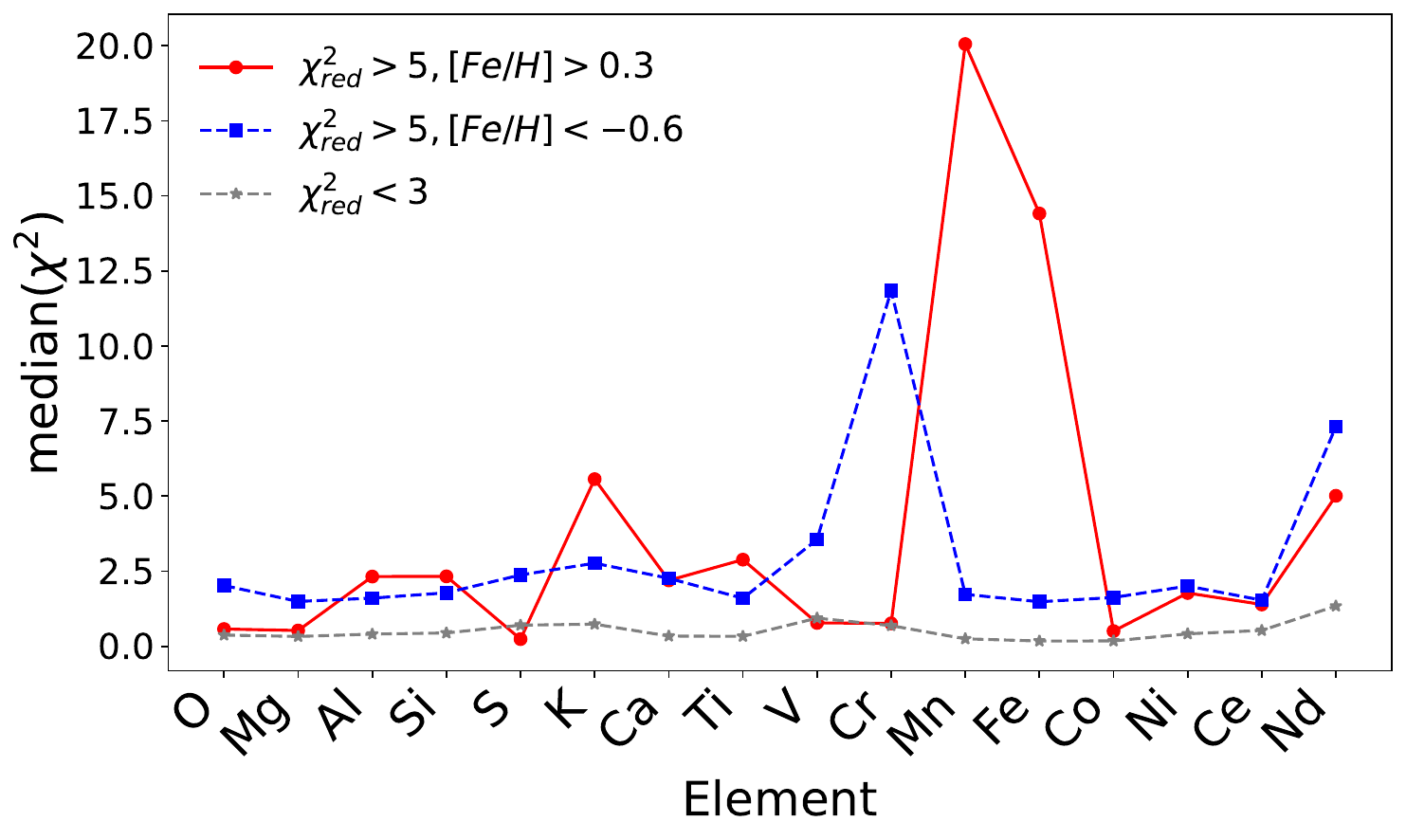}
    \caption{The median per-element $\chi^2$ for the 80\% of stars that have a reduced $\chi^2 < 3$ (grey dashed line) compared to the small fraction (5\%) of stars with reduced $\chi^2 > $ 5 at [Fe/H] $>$ 0.3 ($\approx$ 30\% of stars with [Fe/H] $>$ 0.3) and [Fe/H] < -0.6 ($\approx$ 30\% of stars [Fe/H] $<$ -0.6). Different elements are poorly fit at each metallicity end of the distribution.}
    \label{fig:indiv}
\end{figure}

Finally we examine the per-element failure of stars with a poor overall goodness of fit metric. Figure \ref{fig:indiv} shows the median individual $\chi^2$ of each element for stars considered to be well-fit with $\chi_{\mathrm{red}}^2 < 3$ (80\% of stars), as well as the total $\sim$1\% of poorly-fit stars with  $\chi_{\mathrm{red}}^2 > 5$ and [Fe/H] $> 0.3$ and [Fe/H] $< -0.6$. This figure shows that different elements fail more dramatically -- or are more discriminative in terms of abundance patterns -- at the metal-rich compared to the metal-poor tail of the distribution. Elements K, Mn and Fe show discriminating power away from the latent basis at high metallicity and elements Cr and Nd show the highest discriminating power at low metallicity. While the model failures are valuable for identifying stars with potentially differing enrichment histories to the majority, for the present analysis we exclude these outliers and proceed with the 95\% of stars that are well fit ($\chi^2_{\rm red}<5$).

\begin{figure}
    \centering
    \includegraphics[width=1\linewidth]{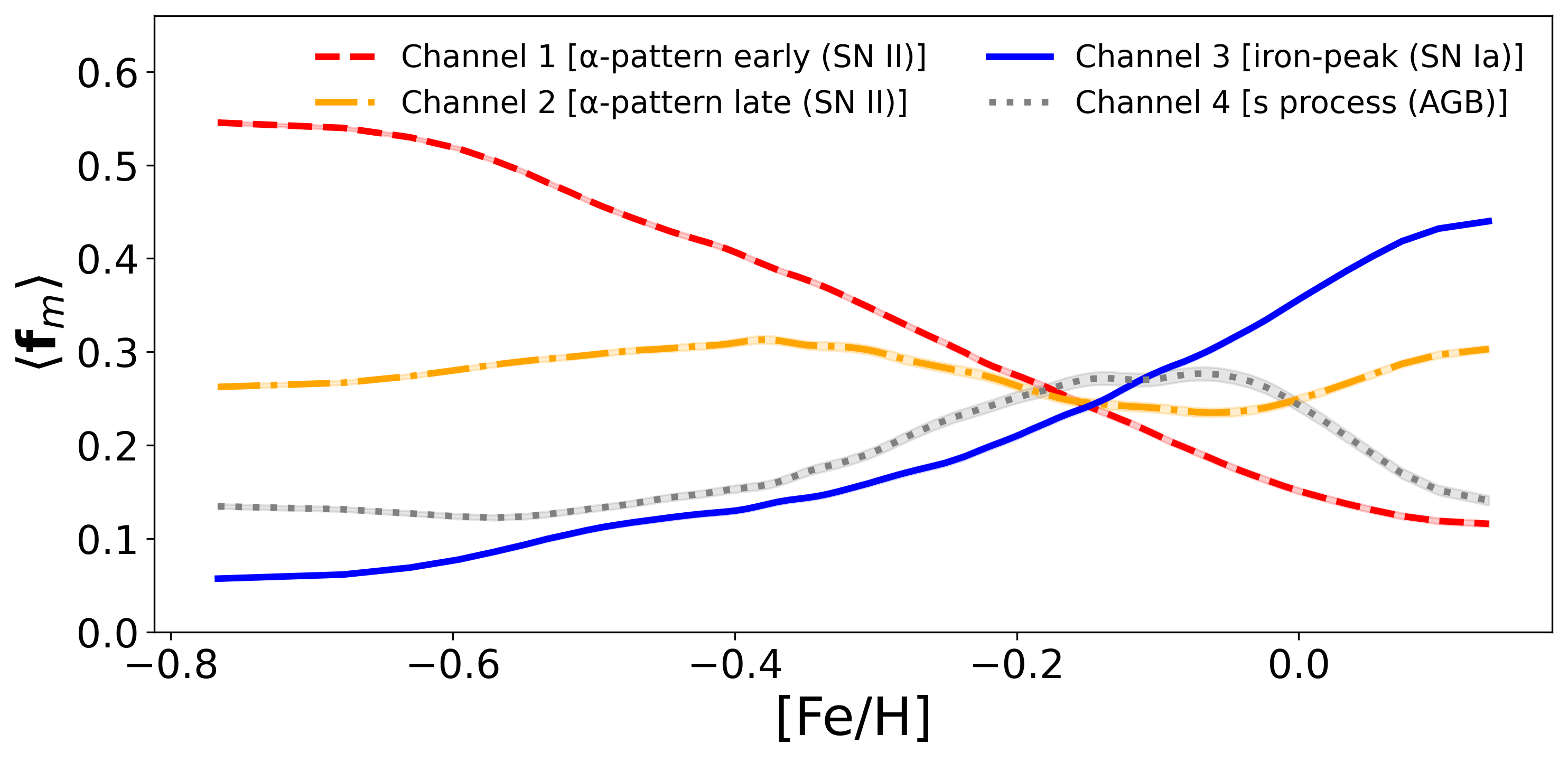}
        \includegraphics[width=1\linewidth]{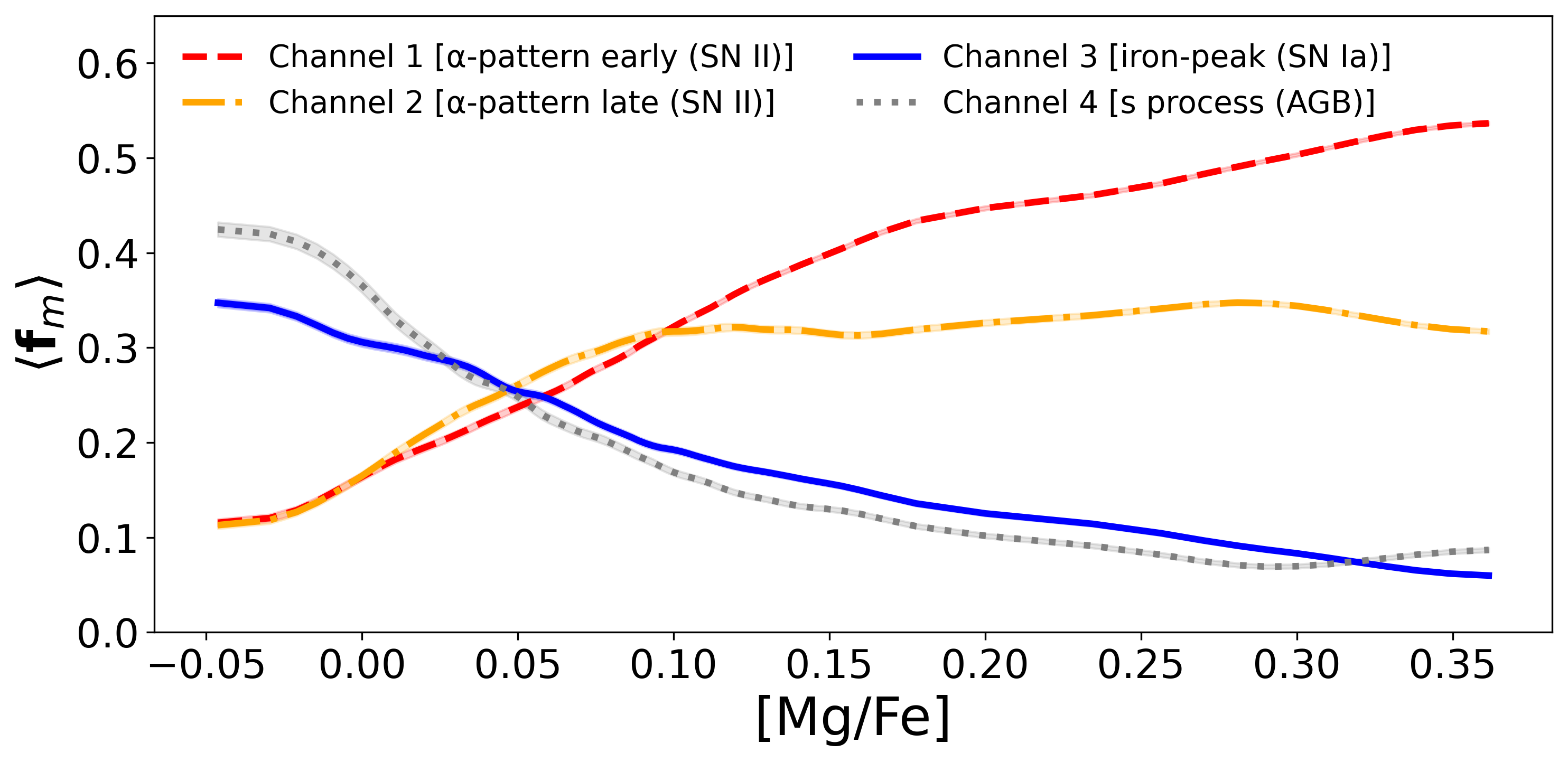}
\includegraphics[width=1\linewidth]{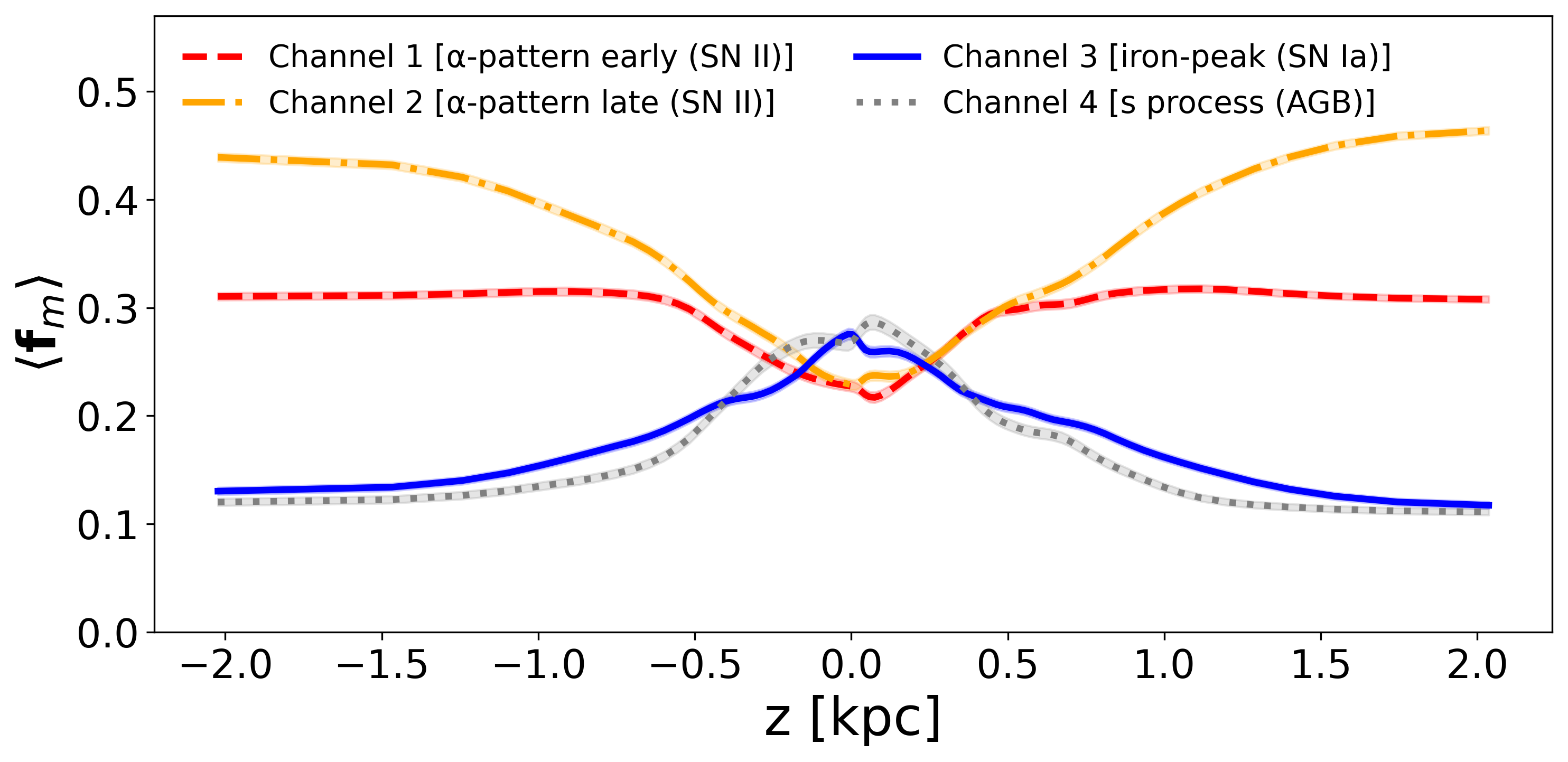}
\includegraphics[width=1\linewidth]{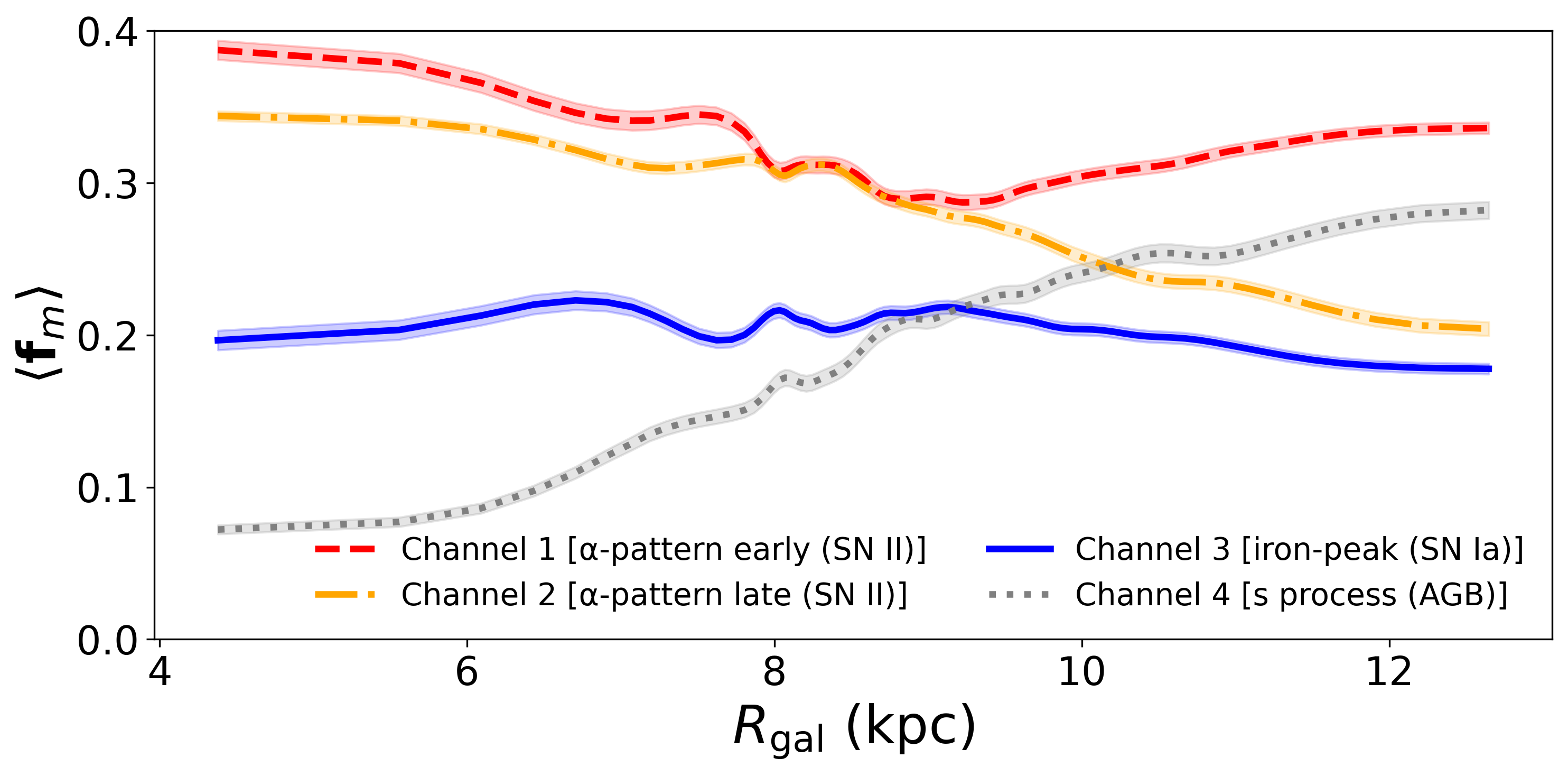}
\includegraphics[width=1\linewidth]{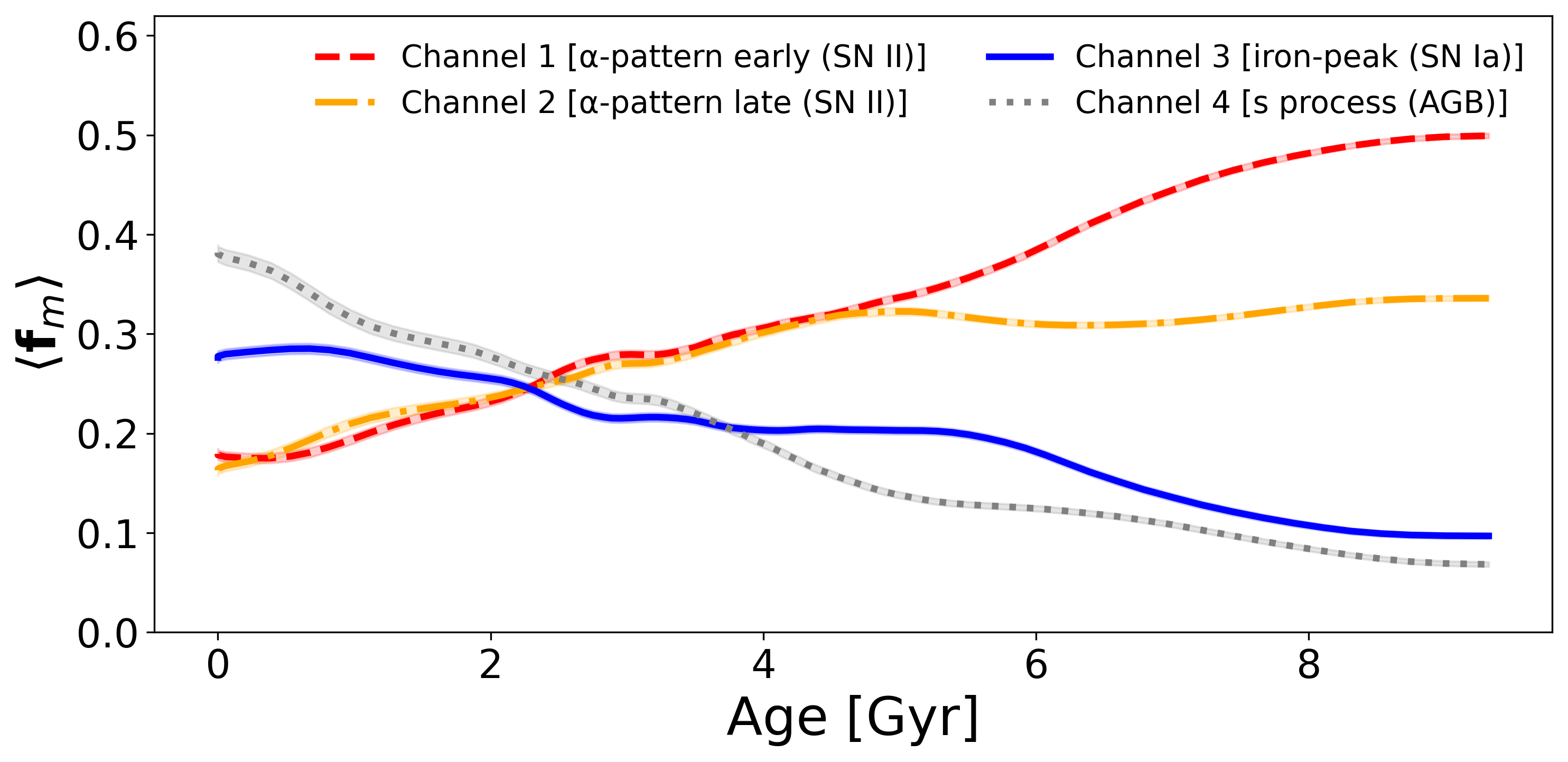}
\caption{Mean per-star normalised pattern fractions, $\langle \mathbf{f}_m \rangle$, shown as a function of five key parameters: metallicity ([Fe/H]), $\alpha$-enhancement ([Mg/Fe]), vertical height from the Galactic plane ($|z|$), Galactocentric radius ($R_\mathrm{{gal}}$), and stellar age. 
Stars are sorted by the relevant coordinate and grouped into bins containing an equal number of stars. All panels show smooth, systematic variations. Channel~1 (SN~II) contributes most strongly at old ages and low metallicity. Channel~4 (AGB) increases toward younger ages and higher metallicity. Channel~3 (SN~Ia) rises sharply at high [Fe/H], coinciding with a minimum in Channel~4. Channels 3/4 enrichment peak near the Galactic mid-plane, and channels 1/2 have  their minimum relative contributions there.}
    \label{fig:fehall}
\end{figure}

\subsection{The Changing Contribution of Enrichment Patterns}

To compare the relative contributions of the latent enrichment patterns across
the Galaxy, we convert the non-negative pattern coefficients (or weights)
$f_{im}$ into per-star normalised fractions. For each star $i$, the vector of
pattern weights is normalised by its sum across all $M$ latent patterns:
\begin{equation}
\mathbf{f}_{im}
=
\frac{f_{im}}{\sum_{m'=1}^{M} f_{im'}} ,
\label{eqn:seven}
\end{equation}
where $m$ indexes the enrichment pattern and the summation runs over all patterns
for a fixed star $i$. By construction,
\begin{equation}
\sum_{m=1}^{M} \mathbf{f}_{im} = 1 .
\end{equation}

These normalised fractions $\mathbf{f}_{im}$ represent the relative contribution
of each latent pattern to an individual star and provide a convenient
parametrisation for examining how enrichment patterns vary with Galactic
position, age, and orbital properties. All fractional pattern maps and trends
presented in subsequent figures use these per-star normalised fractions.

\subsubsection{Fractional Contribution Evolution Across Key Parameters}

Figure~\ref{fig:fehall} shows the mean fractional contribution of each latent enrichment channel for the $\approx$ 65,580 stars with reduced $\chi^2 < 5$, as a function of five key astrophysical parameters: metallicity ([Fe/H]), $\alpha$-enhancement ([Mg/Fe]), vertical height above the Galactic plane ($|z|$), Galactocentric radius, and stellar age. Stars are grouped into bins of 800, such that the physical extent of each bin varies with the local stellar density. Within each bin, we compute the mean of the per-star normalised pattern fractions. We apply a light Gaussian smoothing ($\sigma = 2$ bins) to suppress small-scale fluctuations. Each panel illustrates how the four enrichment channels contribute across these dimensions, averaged over all stars in each bin.

Notably, Figure~\ref{fig:fehall} shows accelerated rates of change for a number of the channels across different variables. This behavior reflects the fact that the decomposition has identified physically coherent abundance patterns that trace changes in the underlying stellar populations and enrichment histories. These are overlapping chemical fingerprints that vary with Galactic structure and time. Across most panels, latent channels~1 and 2, and 3 and 4,  tend to co-vary coherently as a function of each parameters, with the exception of [Fe/H] where channels 3 and 4 diverge at increasing [Fe/H].

The pattern distribution across metallicity, [Fe/H] shows that  channel 1 (SN II; early), is highest at low metallicities, followed by channel 2 (SN II; late), consistent with early, rapid star formation dominated by core-collapse supernovae. As [Fe/H] increases, the contribution from channel~3 (blue; iron-peak from SN~Ia) rises steeply, peaking above solar metallicity, reflecting delayed enrichment by Type~Ia supernovae. Across $[\mathrm{Fe}/\mathrm{H}]$, the two low-mass progenitor channels 3 and 4 are most divergent; channel~3 (SN~Ia) rises while channel~4 (AGB) declines. This may be inherited from different delay-time and metallicity dependancies. 

 As a function of $[\mathrm{Mg}/\mathrm{Fe}]$, Channels~1 and 2 rise and Channels~3 and 4 fall, with a crossover near $[\mathrm{Mg}/\mathrm{Fe}]\sim 0.05$. There is also, notably, an accelerated rate of change in the trajectories of the fractions at $[\mathrm{Mg}/\mathrm{Fe}]\approx 0.2$. Overall, the contribution of channel 1 is about 50\% for [Mg/Fe] $> 0.2$ and at lower [Mg/Fe] it decreases to around 0.1, as does channel 3, while the delayed sources of channels 3 and 4 show the opposite behaviour.

The pattern fractions across height from the mid-plane of the disc, |z| show that the relative contribution of channels 1 and 2 are a mimimum in the mid-plane, and increase to $|z| \sim 1$~kpc and flatten. While the SN~II channels dominate off-plane,  the SN~Ia and AGB channels are highest in the mid-plane and decrease with increasing vertical height; the AGB channel shows a flatter plateau with $z$ relative to SN~Ia, demonstrating the vertical spatial impact of the different progenitor enrichment. The changing pattern fractions across Galactocentric radius shows that the largest gradient across radius is seen for Channel~4 (AGB), which increases with $R_{\mathrm{gal}}$. Channel 3 is reasonably flat, and channels 1 and 2 show a slight decrease across radius.

The patterns across age highlight a clear temporal evolution. Channels 1 and 2 are dominant for the oldest ages ($>6$~Gyr), consistent with early SN~II-driven enrichment. Channels 3 (SN~Ia) and 4 (AGB) both increase toward younger ages, indicating the growing importance of delayed enrichment channels in more recent Galactic history. The channels cross over around an age of $\approx3$~Gyr. All channels are relatively flat in enrichment fraction for stars $> 8$ Gyr and change more rapidly for ages $< 8$~Gyr. The evolution of the channel fractions with age reflects the characteristic timescales of the underlying enrichment sources. However, this does not directly measure their delay-time distributions, as it represents a convolution with the star formation history.

Together, these panels show that the latent channels represent smoothly varying contributions that encode the chemical evolution of the Milky Way. The continuous nature of the transitions reflects that stars form from an evolving mixture of enrichment sources that imprint shared fingerprints that vary in fraction over time, space and metallicity.  This NMF decomposition therefore enables a compact and interpretable representation of that evolution, separating overlapping enrichment histories more directly than traditional abundance–abundance plots.

\subsection{Spatial and Abundance Distributions of Latent Channels}

\begin{figure*}
    \centering
\includegraphics[width=1\linewidth]{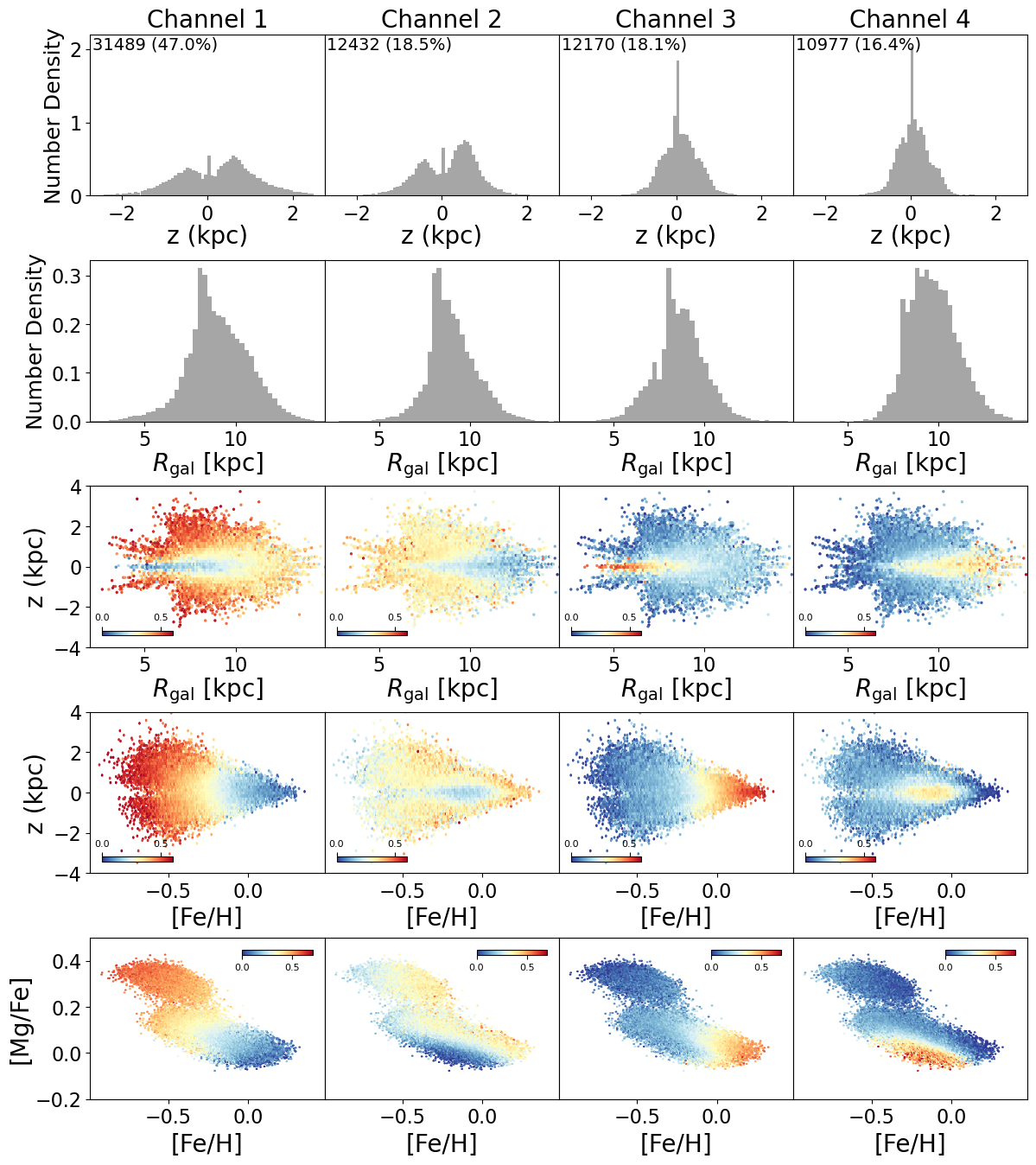}
\caption{
The distinct spatial and chemical distributions of pattern fractions for the $\sim$90\% of stars with reduced $\chi^2 < 5$. The top two rows show the distribution of stars for which each latent channel provides the dominant contribution. The third row shows a spatial map of stars in the $R_{\rm gal}$–$z$ plane, coloured by the per-star normalised pattern fraction $\mathbf{f}_{im}$ for the corresponding channel. The fourth row shows the distribution of pattern fractions across the $[\mathrm{Fe/H}]$–$z$ plane, while the bottom row shows the distribution across the $[\mathrm{Fe/H}]$–$[\mathrm{Mg/Fe}]$ plane.}
    \label{fig:spatial}
\end{figure*}

Figure~\ref{fig:spatial} shows how the four latent enrichment channels are distributed across spatial and chemical dimensions, in histograms and hexbin maps. The top histograms illustrate the vertical ($z$) and radial ($R_{\rm gal}$) distributions of stars where each respective channel is dominant. The lower panels map channel fractions for all stars in two-dimensional planes of $R_{\rm gal}$–$z$, [Fe/H]–$z$, and [Fe/H]–[Mg/Fe].

Channels~1 and 2 (SN~II) are the dominant channels in $\sim$60\% of the stars. These contributions are highest off the disc mid-plane and at small and intermediate radii and at [Fe/H] < -0.2.  Channels 3 and 4 (SN~Ia, AGB) are dominant in about 40\% of the stars and show highest contribution fraction in the mid-plane, in the inner and outer Galaxy, respectively. The maps demonstrate that the latent decomposition has separated physically meaningful abundance patterns that track distinct spatial and chemical trends across the Milky Way. Channels~1 and 2 represent the dominant, early, $\alpha$-enhanced pathways, while channels~3 and 4 reflect delayed enrichment processes that vary systematically with metallicity, radius, and height above the plane.

\begin{figure*}
    \centering
  \includegraphics[width=1\linewidth]{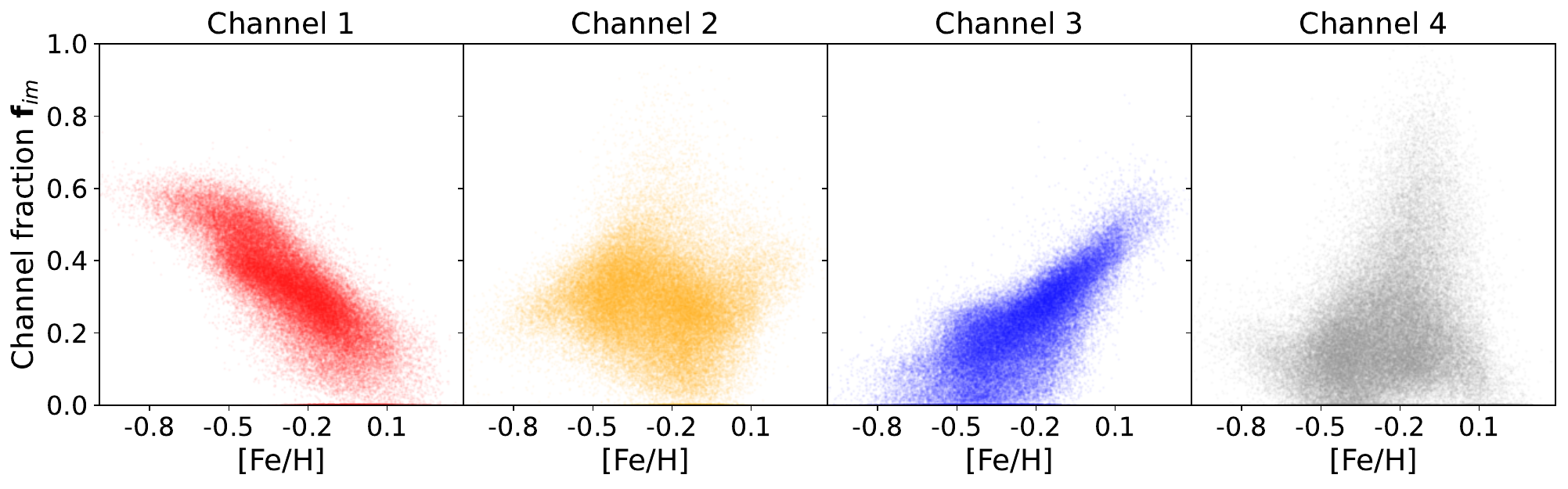}
  \includegraphics[width=1\linewidth]{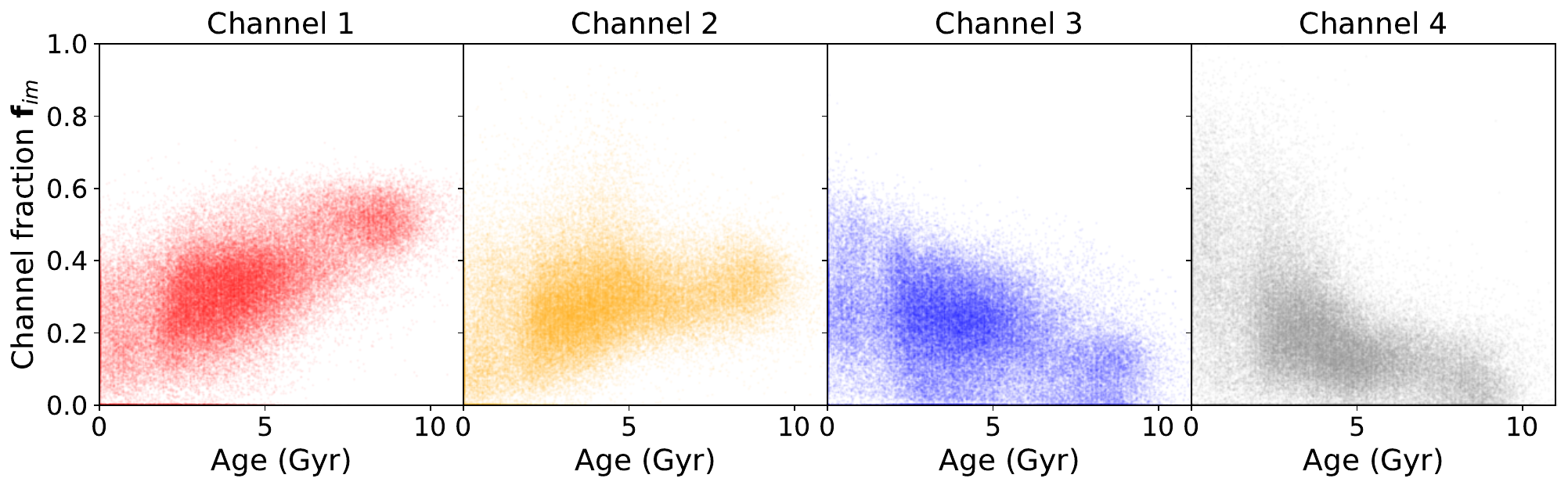}
\includegraphics[width=1\linewidth]{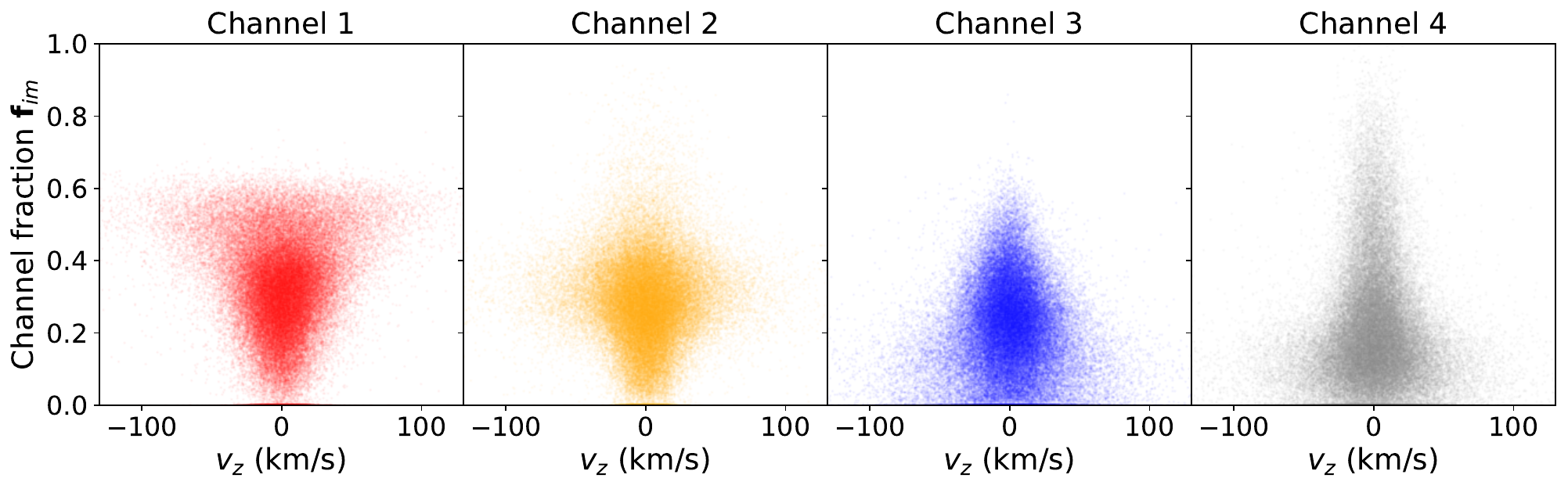}
\caption{ The per-star normalised channel fractions $\mathbf{f}_{im}$ shown on the y-axis for the 90\% of stars with reduced $\chi^2 < 5$.
Each panel shows how the fractional contribution of each enrichment channel (1–4) varies across stellar age, metallicity ([Fe/H]), and vertical velocity ($v_z$). The distributions demonstrate that each channel exhibits a distinct dependence on chemical, temporal, and dynamical properties.}
    \label{fig:allchan}
\end{figure*}

Figure~\ref{fig:allchan} shows the per-star fractional contributions of each latent channel, plotted against [Fe/H], age, and vertical velocity $v_z$. Whereas the Figure \ref{fig:spatial} showed binned averages across spatial and abundance dimensions, and Figure \ref{fig:feh_fraction} showed binned averages per channel, here we see the full distribution of stars and the spread within each channel across key variables.

Similarly to Figure \ref{fig:spatial}, clear differences emerge. Channel~1 (SN~II; early) shows the strongest positive correlation with age and extends to the broadest $v_z$ distribution, consistent with an older, dynamically heated population. Channel~3 (SN~Ia) is strongly correlated with [Fe/H], rising steeply toward solar metallicities. 

This star-by-star view highlights that the channels not only trace distinct mean trends but also differ in their scatter and dynamical imprint. In particular, the broader $v_z$ spread of Channel~1 and the metallicity sensitivity of channel~3 demonstrate that the decomposition isolates chemically and dynamically distinct pathways, beyond what is visible from mean fraction trends alone.

Altogether, the correlations between the fractional contributions of the enrichment channels and the spatial, kinematic, chemical, and temporal properties of the stars reveal how the expression of each pattern varies across the disc. The patterns trace different nucleosynthetic pathways, and their spatial and dynamical behaviours are consistent with the sources they represent. Each channel dominates in different regions of the disc and at different epochs, and exhibits distinct kinematic signatures that reflect the integrated star-formation and enrichment history of the Milky Way.

\begin{figure*}
    \centering
\includegraphics[width=1\linewidth]{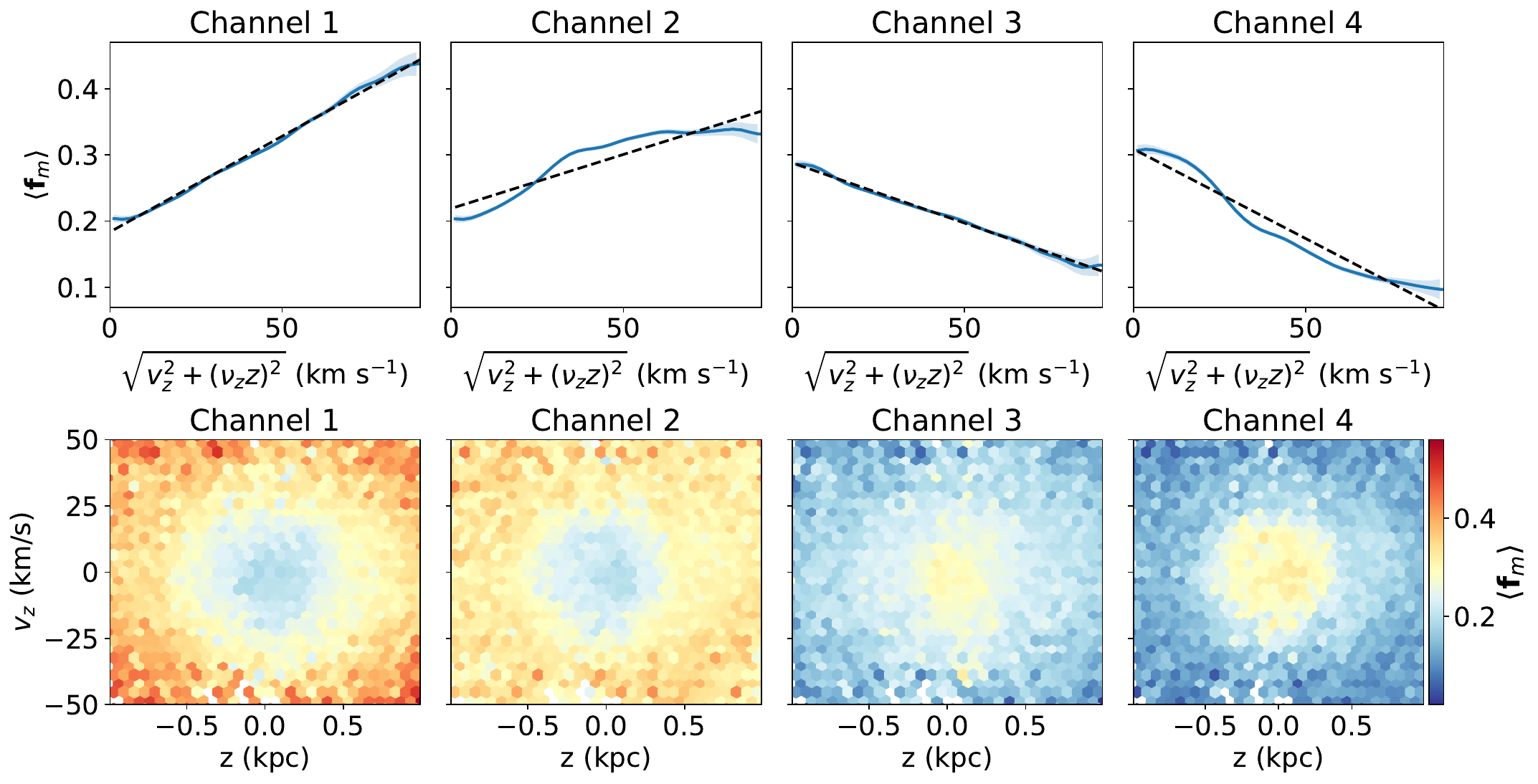}
 \caption{The bottom row shows the distribution of stars across the $z-v_z$ plane for each latent channel, coloured by their pattern fractions $(f_{ch})$. The location in the $z-v_z$ plane is a proxy for vertical orbital energy, which increases with distance from the center of these coordinates. The fractional distributions of latent channel shows structure that varies between the channels with opposite contrast between channels 1/2 and 3/4. Channels 1 and 2 (early and late SN II-associated, respectively) show higher contributions at larger orbital energies and channels 3 and 4 (SN~Ia and AGB associated, respectively) show higher concentrations at low orbital energies. The top row of panels shows the mean channel fraction, $\langle \mathbf{f_{m}} \rangle$, as a function of the vertical energy proxy, $x \equiv \sqrt{v_z^2 + (\nu_z z)^2}$. The solid blue lines show the smoothed mean values of $\mathbf{\langle f_m \rangle}$, with shaded regions indicating the standard error. The dashed lines represent a best-fit linear model. Deviations of the data from this simple model is seen in Channels 2 and 4, indicating that these channels are not fully described by a simple energy-dependent relation.}
    \label{fig:snail}
\end{figure*}

\subsection{Dynamical Signatures in Latent Patterns}

In Figure~\ref{fig:snail} (bottom panel) we show the fractional contribution of each latent channel in vertical phase space ($v_z-z$), for the 44,473 stars with $|z|<1$~kpc, Galactocentric radii $6<R_{\rm gal}<11$~kpc, and parallax uncertainties $<20$\%. A central structure is visible in all four channels, though with varying strength and morphology. This provides a chemo-kinematic view of the disc, revealing that stars separate kinematically according to their enrichment pathways \citep[e.g.,][]{Antoja2018, Frankel2025, Horta_oti}.

The top panels of Figure~\ref{fig:snail} show, in the blue lines, the smoothed mean channel fraction as a function of a vertical energy proxy, defined as
$x \equiv \sqrt{v_z^2 + (\nu_z z)^2}$,
where $z$ is the height above the plane, $v_z$ is the vertical velocity, and $\nu_z$ is the vertical frequency of a circular orbit in the midplane evaluated at each star’s Galactocentric radius. Stars at lower values are kinematically ``colder" and more confined to the mid-plane, while those at higher values are ``hotter” and reach larger vertical excursions. The gradients of the mean channel fraction are therefore diagnostic of the orbital behaviour associated with each channel. Channels~3 (SN~Ia) and 4 (AGB) show negative gradients: their contributions are highest for dynamically cold stars and fall off at higher orbital energies. By contrast, channels~1 and 2 (early and late SN~II) show positive gradients, with their contributions rising at higher orbital energies, consistent with these tracing old, dynamically hot populations.

The dashed lines in the top panels of Figure \ref{fig:snail} show a linear fit to the mean channel fraction as a function of the vertical energy proxy $x$, defined as $ \langle f_m \rangle = a + b\,x$, where $x = \sqrt{v_z^2 + (\nu_z z)^2}$, with $a$ and $b$ obtained from a linear regression using the \texttt{scipy.stats.linregress} function.  The systematic deviations of the running mean of the data from this simple model are informative and show that the enrichment channels retain distinct dynamical imprints in the detail of their phase-space gradients. Channels 1 and 3 (early SN II and SN Ia, respectively) appear to follow the smooth trend more closely, and channels 2 and 4 (late SN~II and AGB, respectively) show larger undulating deviations of the data from the simple model description. This is not a trivial consequence of the fractions summing to unity, but evidence for a genuine dynamical difference in the stellar populations. The stronger deviations in channels 2 and 4 are consistent with these components tracing more dynamically mixed populations, and may also reflect sensitivity to phase-space structure from dynamical perturbations.

\section{Discussion}
\label{discussion}

This study introduces a data-driven, generative framework that uses  non-negative matrix factorization (NMF) to model stellar abundances as  combinations of a small number of shared latent patterns. The motivation  follows earlier work suggesting that abundance distributions in the Milky Way  disc can be represented as fractional mixtures of a few nucleosynthetic  sources. Our goal is to examine the Galaxy using dominant signatures of its enrichment history encoded in the joint abundances.  To do so, we rely not on the absolute abundance scale of each star, but on the  \textit{shape of its abundance pattern}—the relative differences between  elements. It is this pattern structure that we want to access, as the encoding of the different nucleosynthetic pathways underlying each star. Accordingly, our  NMF decomposition is designed to emphasise pattern shape.  The enforcement of positivity via a per-star abundance offset is conceptually analogous to continuum normalisation in stellar spectroscopy, where a global scale is removed to emphasise relative structure, for which simple data-driven representations are particularly effective \citep{Ness2015}. In this formulation,  each star’s abundance vector is expressed as an additive combination of shared  patterns, each of which encodes a characteristic abundance shape. In the abundance context, this yields a linear and interpretable representation in which the large vector of abundances data are organised into a compact set of channels, with each star described by its fractional contributions to them.

Although mathematically sound, the NMF decomposition must be interpreted with caution. NMF can capture not only nucleosynthetic signatures but also survey systematics, sample-selection effects, and correlated evolutionary trends that appear in the data. The latent patterns therefore represent the statistically dominant modes of variance in the survey, with a nucleosynthetic interpretation becoming meaningful only when supported by astrophysical expectation and independent behaviour across time, space, and dynamical properties. While the pattern fractions do show validating structure across independent parameters, latent patterns should not be interpreted as direct stellar yields or absolute source fractions (i.e., absolute measurements of SN~II/SN~Ia and AGB contributions). Instead, they reflect the integrated chemical evolution of the disc and the dominant modes of element production present in the data, that can be reasonably connected to specific sources.

The framework is similar in spirit to the latent variable model of \citet{Casey2019}, in which the data are used to simultaneously infer nucleosynthetic patterns. However, our model is generative, enabling us to validate the decomposition directly, and identify regions where the model fails and to highlight populations whose chemical enrichment histories depart from the dominant trends. These deviations may correspond to  distinct events, rare enrichment channels, or accreted origins.  The approach here also shares conceptual ground with the $k$-process model of \citet{Weinberg2022,Griffith2024}, which models elemental production as a mixture of physically motivated nucleosynthetic sources.  In contrast, the NMF decomposition used here is fully data driven, does not assume any a priori mapping between elements and processes, and imposes no restricted correlation structure between abundances and sources.  The resulting model is mathematically simpler and solved via a linear, non-negative factorisation. This provides a direct empirical description of the enrichment manifold. As the NMF model is a low-rank matrix factorisation, a separation of the population into a train and test sample is less critical than in regression models with many degrees of freedom. Here, all stars are constrained to be factorised into a common low-dimensional subspace. Nevertheless, we repeated the analysis using a 50/50 train--test split, learning the enrichment patterns on one subset of stars and applying them to the remainder. The reconstruction statistics change very slightly, but the overall behaviour is unchanged, indicating that the low-dimensional structure is stable and not driven by overfitting.

\noindent Our main findings are summarised as follows:

\begin{itemize}
    \item We find that $M=4$ latent abundance patterns are sufficient to reconstruct the 16-element abundance vectors of most stars, with good generative model performance in the range $-0.6 <$ [Fe/H] $< 0.3$. Model performance declines rapidly outside this range.  Approximately 5\% of stars overall are poorly reproduced by the model. Model failure, where enrichment pathways diverge, is most substantially expressed in Cr at low-[Fe/H] and Mn at high-[Fe/H]. At low metallicity, [Fe/H] $< -0.6$, the disc appears to be an increasingly mixed population from multiple enrichment pathways, including accreted ex-situ stars that are dynamically distinct, and stars in unremarkable orbital phase-space that the model cannot capture with $M=4$.   Even up to $M=8$, model failures persist, albeit at a lower rate. This model failure for metal poor stars is indicative of the underlying multitude of enrichment channels with distinct patterns and gas inflows from external systems. This finding is consistent with \citep{Jo} who find that the last massive merger contributed to the chemical enrichment and building of the metal-rich part of the thick disc at early times.

    \item We associate the high-precision latent abundance patterns ($\sigma_P\sim3$\%) with imprints of massive core-collapse supernovae (SN~II), with a separation of this channel into early and late contributions, low-mass supernovae (SN Ia), and Asymptotic Giant Branch (AGB) enrichment. The different patterns of nucleosynthesis dominate in different spatial regions and have different kinematic signatures and timescales, showing that the relevant enrichment modes change over time and spatial location in an organized way. Even without invoking specific nucleosynthetic associations, the model clearly yields a low-dimensional representation of the abundances that decomposes the data into components with distinct spatial, chemical, age, and orbital behaviours, thereby linking naturally to the disc’s formation and evolutionary history.

\end{itemize}

\subsection{Implications of a Low dimensional Basis}

Galactic chemical evolution is governed by nonlinear processes. This includes gas flows, star formation, turbulence, and feedback. Despite this complexity, we find, consistent with prior work, that a simple low-rank linear model in log(abundance) reproduces the observed abundance patterns with high fidelity.

This emergence of low-dimensional structure from a system governed by nonlinear physics is not unique to Galactic chemical evolution. Similar behaviour is observed in other complex systems. For example, gene expression studies using NMF show that thousands of gene expression measurements can be represented as linear combinations of a very small number of shared patterns (or ``metagenes", \citet[e.g][]{Brunet2004}). More generally, nonlinear dynamical systems can be well described by low-dimensional representations when the dynamics are governed by a limited number of dominant processes \citep{Brunton2016}.

In the context of the Milky Way disc, the non-linear physics does not imply high-dimensional observables. Rather, the generative nature of a low-dimensional linear model must mean that the number of independent degrees of freedom driving the primary variation is small. Although many physical processes contribute to enrichment, they are not independent, and instead are coupled through shared drivers such as star formation history, gas supply, and delay-time distributions (see Zhang et al., in preparation). The observed abundance space is therefore low-dimensional, and a linear model provides an efficient description of how these dominant enrichment patterns combine across stars.

We emphasise that this model describes the ensemble structure of the data rather than the underlying time evolution and generative physics, which remains nonlinear. The model does not perfectly reproduce all abundances: the residuals contain additional structure likely arising from a combination of  systematics, and possibly genuine astrophysical complexity beyond the dominant enrichment channels (e.g. metallicity-dependent yields, multi-zone mixing, or rare enrichment events). These deviations ultimately suggest a hierarchical description of Galactic chemical evolution, in which a small number of dominant processes set the bulk abundance patterns, while secondary processes may introduce higher-order structure \citep[e.g.,][]{Griffith2025}.

\subsection{Universal Evolution of the disc}

We find a shared basis of patterns that describe the abundance patterns of disc stars; the fractional contribution of these patterns changes in a structured way across time, spatial dimensions and metallicity. Generative model failures, particularly at [Fe/H] $< -0.6$ indicate increasing diversity in enrichment pathways at low metallicity in accreted stars and gas. This highlights the possible role of merger driven gas inflow at the low-metallicity end of the disc \citep{Kh2021, Buck2020, Renaud2021, Agertz2021, Hanna2025}. Where the model is successful, however, the observed smooth trends in the latent space and the continuity of enrichment traced by both the latent channel fractions and their trajectories are consistent with evolving gas inflow modulating the pace of enrichment.

We can connect these results directly to the observed $\alpha$-bimodality in the Galactic disc. This is a visually striking valley in the stellar number counts across $\alpha$-elements with respect to iron, which occurs around [Mg/Fe]$\sim$0.2 in the Milky Way Mapper data.  Our analysis reveals that the high- and low-$\alpha$ disc populations are modeled by the same underlying enrichment channels, but in differing proportions. We see a changing relative emphasis of enrichment sources over time and over [Mg/Fe] in the mean fractional contributions shown in Figure~\ref{fig:fehall}. At high [Mg/Fe], the population is dominated by the SN~II channels, each contributing $\approx 35-50$\% of the enrichment. However, once [Mg/Fe] falls below $\sim$0.2, the SN~II contribution declines steeply, with channels 1 and 2 dropping to $\approx$10\% each by [Mg/Fe] $\approx -0.05$. In parallel, the delayed nucleosynthetic sources, the SN~Ia and AGB channels, rise sharply in fractional contribution. This accelerated handover from prompt to delayed enrichment helps explain the observed paucity of stars with intermediate [Mg/Fe]: the interstellar medium evolved rapidly through this regime, leaving relatively few stars formed during the transition. 

In this framework, we interpret the origin of the $\alpha$-bimodality as arising from a shift in the relative dominance of enrichment channels, rather than from two fundamentally distinct gas reservoirs. The continuity of the recovered latent abundance patterns across the $\alpha$-valley indicates that the bulk of the Galactic disc is well described by a shared, low-dimensional chemical basis. This is naturally aligned with a chemically self-regulated evolutionary system.
This interpretation relies on the assumption that stars formed within a common evolutionary system share a characteristic abundance basis.
Distinct systems with sufficiently different enrichment histories would not, under this assumption, be expected to project onto the same latent patterns. Consistent with this expectation, we do see that stars for which the model fails to reproduce the observed abundances are not randomly distributed, but instead preferentially occupy regions of phase space associated with accreted material and the metal-poor disc.
These failures therefore provide complementary evidence that the dominant latent basis captures the in-situ disc evolution, while deviations highlight populations with distinct origins.

Simulations show that such transitions in chemical enrichment can result from gas accretion from the circumgalactic medium \citep{Hanna2025}, including both ambient gas and gas-rich mergers. While stars not well fit by our model at [Fe/H] $< -0.6$, which are the majority at this [Fe/H], are consistent with high-$\alpha$ stars formed during such early, merger-driven gas inflow, the bulk of the stellar population, including both the high- and low-$\alpha$ sequences, is fit under a single modeling framework. This implies that while both ambient and merger-driven gas accretion are at play in the disc's formation, the disc evolution at [Fe/H] $> -0.6$, and across the $\alpha$-bimodality, is consistent with regulated gas accretion from the circumgalactic medium.

\subsection{Model Failures as an Opportunity and Future Prospects}

Model failures, whereby a star's abundances are not well-generated by the latent basis, can arise for two reasons. Such failures may reflect stars with different astrophysical enrichment histories which have dissimilar abundance pattern structure to the majority of the stars. Alternatively, model failures may  indicate where the simple linear model does not fully capture the pattern diversity of the disc. Because the model reconstructs the abundances for the majority of stars, the low-dimensional structure is clearly present in the data. The failures simply highlight where stars deviate from it. Subsequently, stars that are not well captured by the model represent valuable candidates for future study. This approach is clearly a useful way to isolate such stars In some cases, the stars that fail the model generation show clustering in dynamical space. In other cases, their dynamical properties are unremarkable.  In the case where the dynamics is unremarkable, there may be astrophysical reasons nevertheless that cause a different pattern structure in their abundances. Fitting with additional latent variables beyond $M = 4$, thereby increasing model flexiblity, improves the generative fit for a subset of poorly fit stars. These patterns that need enhanced model flexilibty could have arisen due to mixing of external gas with the existing circumstellar material. This would be consistent with merger-driven gas inflow, supplementing the ambient gas regulated by the disc’s internal processes \citep[e.g.,][]{Hanna2025}. Some poorly fit stars, on the other hand, may have formed ex-situ, particularly those with high eccentricities and low angular momentum seen in Figure \ref{fig:outliers}. 

Comparison with physically motivated methods \citep[e.g., the k-process Model;][]{Griffith2024} that solve for fractional contributions of assigned sources, given the abundances, highlights complementary strengths, where forward models embed theoretical priors. Our empirical framework offers a new means to constrain the timing and nature of key enrichment transitions, and may help distinguish between episodic and continuous formation scenarios when compared to forward chemical evolution models.

This approach can be applied more generally across different systems to test the universality of low-dimensional chemical bases, for example in dwarf galaxies,
and to connect observational data with simulations in a way that does not depend on detailed agreement in absolute nucleosynthetic yields.

We note that similar results to those showcased here can be obtained with \apogee\ DR17 abundances, although the generated model slightly better matches the measured abundances, as evaluated by the reduced $\chi^2$, using Milky Way Mapper abundance measurements.

\subsection{Model Limitations}

While our NMF-based framework empirically discovers enrichment patterns, offering a flexible and interpretable approach to latent structure in elemental abundances, it also highlights limitations and areas for further development. 

\subsubsection{A non-unique solution}

Results depend on the specific set and characteristics of elemental abundances included in the decomposition, as well as how abundances are input and normalised, since NMF groups stars based on the global structure of the input abundance space. Our modeling is therefore sensitive to the choice of number of latent channels, the number of elements, the normalisation of the elements, and the weighting used to handle measurement uncertainties. Although latent models can recover statistical correlations, we argue that the shared latent patterns we find, whose fractional contributions vary systematically across the disc, do reflect the combined influence of stellar yields, star formation history, and chemical mixing, dominated by the different associated enrichment sources. These show correlation across age, spatial extent and kinematics that connect to the enrichment channels we associate the patterns with.

Despite the non-unique nature of our solution, we find that the primary conclusions of this work are robust. Different choices enable the same low-basis factorisation that shows patterns with fractions that vary spatially and dynamically. The consistent recovery of dominant channels under tests of different elements, number of latent variables and uncertainty weighting schemes supports the interpretability of the latent decomposition. Nonetheless, we emphasise that the solution is not unique, and latent space should be, in general, interpreted with caution. Furthermore, while our interpretation of the fractional changes in latent channel contributions assumes that the learned patterns correspond to real enrichment sources, our conclusions do not depend on that interpretation.

\subsubsection{Structured residuals}

The structured residuals between the model and the measured abundances, visible in both the [Fe/H]–[Mg/Fe] plane and the spatial maps (see Appendix), indicate the presence of additional processes that systematically influence elemental abundances but are not captured by our current model (e.g., local enrichment, or the need for more flexible non-linear representations).  Notably, much of the residual structure correlates with observational  variables such as SNR and stellar effective temperature.  This suggests that  the latent model is also absorbing systematic artefacts arising from abundance measurements themselves.  A practical advantage of simpler regression-based approaches, which offer different analysis outputs to those here, is that such observational quantities (e.g.\ $T_{\rm eff}$, $\log g$, SNR) can be explicitly included as predictors, allowing their impact on abundance estimates to be modelled and removed \citep{Ness2022, Mead2025}.

\subsection{Future Work}

The patterns recovered here reflect the dominant structure present in the available abundance space, and are therefore expected to be broadly consistent across different datasets, provided a similar set of elements is measured at comparable precision. Expanding the element set, for example through the inclusion of additional neutron-capture elements, introduces new independent constraints that can refine the decomposition and may reveal additional structure. With additional heavy elements, a larger number of patterns may be required to adequately describe the data. This work establishes a reference framework for the factorisation of stellar abundances. The approach can be applied across a wide range of datasets, from small, high-fidelity samples of a few hundred stars to large spectroscopic surveys, including those with broader elemental coverage such as GALAH \citep{Buder2025}. A full recovery of the underlying nucleosynthetic processes, independent of the dataset, will require both expanded element coverage and validation on controlled synthetic datasets where the true enrichment channels are known. Therefore, an avenue for future work will be to assess this decomposition approach across datasets, and to validate them in simulations where the true enrichment history is known. Using simulations, this approach may offer a path to establish a mapping between observed enrichment patterns and the generating star formation history.

\section{Conclusion}
\label{conclusion}

Our goal is to develop an approach to decode the joint information that is expressed in the shape of stellar abundance patterns. Our analysis deconstructs (shifted, to be non-negative) stellar abundances into a set of shared enrichment patterns, with varying contributions for each star.  This provides a new modeling approach for understanding the Milky Way disc that links the ensemble of abundance measurements directly to enrichment history, rather than relying on individual elements as isolated tracers. 

The latent enrichment patterns serve as meaningful tracers of Galactic evolution, with their contribution fractions correlating with spatial, chemical and kinematic parameters. In this context,  the disc’s history can be interpreted as a transition in enrichment source diversity, from populations dominated by prompt SN~II sources to those shaped by a more heterogeneous combination of nucleosynthetic channels.  The shared chemical basis that describes both high- and low-$\alpha$ stars implies that these populations can form from the same underlying gas reservoir, sampled in varying proportions as the interstellar medium evolved. Conversely, if the high- and low-$\alpha$ sequences had formed from entirely distinct reservoirs of gas (for example, through gas-rich mergers), each with independent enrichment histories, we would not generally expect the full disc to be well described by a shared set of four latent abundance channels. An exception would be if nucleosynthetic yields themselves collapse onto a low-dimensional basis largely independent of the detailed enrichment environment.

The NMF analysis in this work subsequently reframes the chemical evolution of the Milky Way as a transition in enrichment source diversity. In addition, it establishes a template for interpreting forthcoming data. Although developed here with a high-fidelity subset of Milky Way Mapper data, the method can be applied to lower-quality datasets and scaled to larger samples, including upcoming surveys with expanded elemental coverage \citep{deJong2016, C2020, Gonzalez2020}. A recent application in the stellar halo \citep{Davies2025} demonstrates its broader potential.

When combined with additional tools, this approach may provide a pathway to recover nucleosynthetic yields from abundance data. In addition, the NMF approach offers a path forward for future efforts to model Galactic enrichment and structure jointly. It offers promise for turning to other galactic environments. This includes low-metallicity dwarf galaxies and globular clusters, where it can similarly be used to quantify enrichment patterns and transitions in the populations, and reveal how different enrichment patterns connect to their chemical and dynamical characteristics. As next-generation surveys deliver new samples of stars and elements, this approach provides a route to unify Galactic and extragalactic archaeology under a common, interpretable foundation.

\section*{Data Availability}

The data used here are part of the SDSS-V Milky Way Mapper survey. Milky Way Mapper data are made publicly available through the Sloan Digital Sky Survey data releases (DR19).

\section{Acknowledgements}

We thank the reviewer for their helpful and thoughtful comments that have improved the manuscript.
Thanks to Trevor Mendel, Sven Buder, Luca Casagrande for helpful discussions and Kiyan Tavangar and Adrian Price-Whelan for helfpul discussions with the $z-v_z$ plane in particular. Some of the text and code refinements were developed in micro-managed highly supervised engagement with AI language models (ChatGPT, OpenAI and Gemini, Google).

Funding for the Sloan Digital Sky Survey V has been provided by the Alfred P. Sloan Foundation, the Heising-Simons Foundation, the National Science Foundation, and the Participating Institutions. SDSS acknowledges support and resources from the Center for High-Performance Computing at the University of Utah. SDSS telescopes are located at Apache Point Observatory, funded by the Astrophysical Research Consortium and operated by New Mexico State University, and at Las Campanas Observatory, operated by the Carnegie Institution for Science. The SDSS web site is \url{www.sdss.org}.

SDSS is managed by the Astrophysical Research Consortium for the Participating Institutions of the SDSS Collaboration, including the Carnegie Institution for Science, Chilean National Time Allocation Committee (CNTAC) ratified researchers, Caltech, the Gotham Participation Group, Harvard University, Heidelberg University, The Flatiron Institute, The Johns Hopkins University, L'Ecole polytechnique f\'{e}d\'{e}rale de Lausanne (EPFL), Leibniz-Institut f\"{u}r Astrophysik Potsdam (AIP), Max-Planck-Institut f\"{u}r Astronomie (MPIA Heidelberg), Max-Planck-Institut f\"{u}r Extraterrestrische Physik (MPE), Nanjing University, National Astronomical Observatories of China (NAOC), New Mexico State University, The Ohio State University, Pennsylvania State University, Smithsonian Astrophysical Observatory, Space Telescope Science Institute (STScI), the Stellar Astrophysics Participation Group, Universidad Nacional Aut\'{o}noma de M\'{e}xico, University of Arizona, University of Colorado Boulder, University of Illinois at Urbana-Champaign, University of Toronto, University of Utah, University of Virginia, Yale University, and Yunnan University.

R. L-V. acknowledges support from Secretar\'ia de Ciencia, Humanidades, Tecnolog\'ia e Inovacci\'on (SECIHTI) through a postdoctoral fellowship within the program ``Estancias posdoctorales por M\'exico''

Y.Y.S. acknowledges support from the Dunlap Institute, which is funded through an endowment established by the David Dunlap family and the University of Toronto.

LC thanks the support from grant IA103326 (DGAPA-PAPIIT, UNAM) and project SECIHTI CBF-2025-I-2048.

\section{Appendix}

\subsection{Abundance Residuals}

The element reconstruction using the NMF framework reveals structured residuals, which appear both spatially and across chemical abundance space. These residuals, defined as the difference between measured and NMF-reconstructed abundances, arise in part due to the approximation inherent in the compression of 15 observed abundance dimensions into only $M=4$ latent enrichment channels. This dimensionality reduction limits the ability to fully reconstruct the detailed abundance distribution of every element.

\begin{figure*}
    \centering
        \includegraphics[width=1\linewidth]{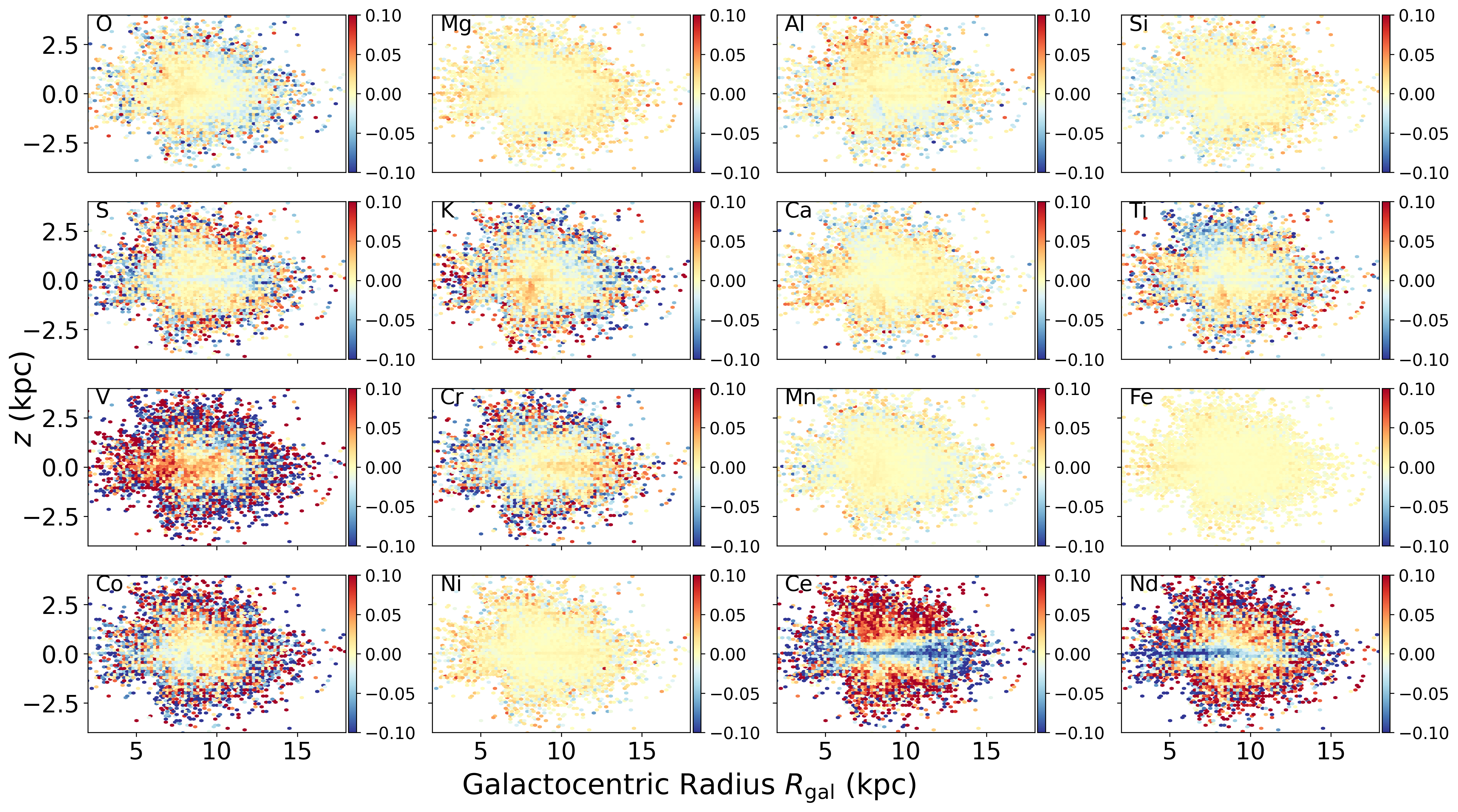}
    \caption{The $R_{\mathrm{gal}}-z$ plane coloured by the mean residuals (model prediction - measurement) of the generated abundances - the measured abundances across the $R-z$ plane for all 15 elements. The colourbar shows the residual amplitude and the element for which the residual is calculated  is shown in each sub-panel.}
    \label{fig:resid}
\end{figure*}

In Figure~\ref{fig:resid}, we show the distribution of the residuals
\((\mathrm{[X/H]_{measured}} - \mathrm{[X/H]_{model}})\) for each element,
where \(\mathrm{[X/H]_{model}}\) is evaluated as in Equation~\ref{eqn:chisquared}: $\mathrm{[X/H]_{model}} = (fP)_{ij} + \min_{j'} X_{ij'} - \epsilon_{ij}^2 .$

Some elements show larger systematic offsets than others, likely reflecting that certain nucleosynthetic signatures are not fully captured by the limited latent basis. However, the residuals also show spatial structure that extends beyond what would be expected from an underfit alone.  

\begin{figure*}
    \centering
        \includegraphics[width=1\linewidth]{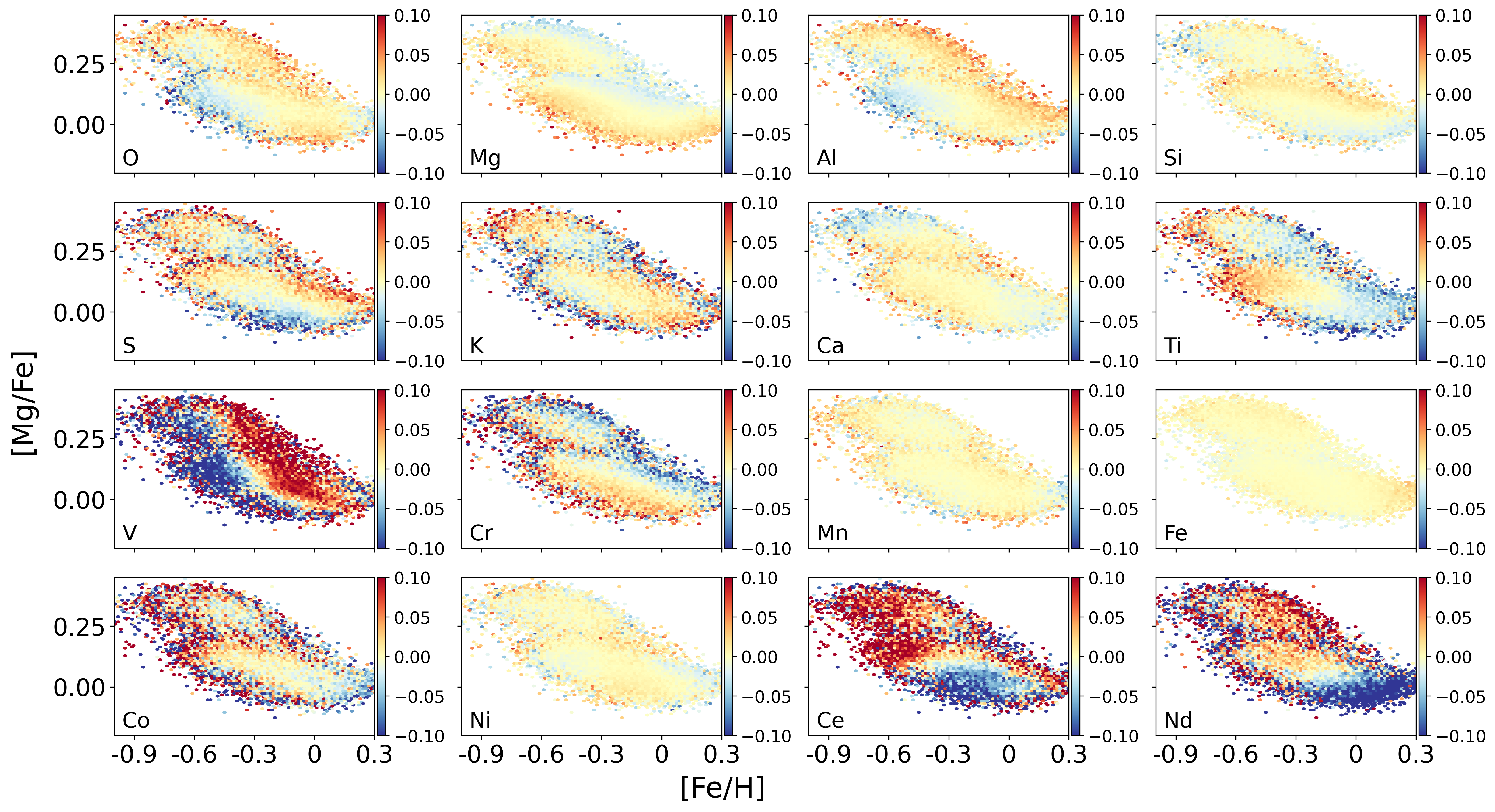}
    \caption{The [Fe/H]-Mg plane coloured by the mean residuals (model prediction - measurement) of the generated abundances - the measured abundances across the $R-z$ plane for all 15 elements. The colourbar shows the residual amplitude and the element for which the residual is calculated  is shown in each sub-panel.}
    \label{fig:resid2}
\end{figure*}

In Figure~\ref{fig:resid2}, we project the residuals onto the \([\mathrm{Fe/H}]\)–\([\mathrm{Mg/Fe}]\) plane. Here, coherent patterns emerge, with specific regions of the abundance plane showing systematically over- or under-predicted values for particular elements. These patterns are not random and suggest missing enrichment pathways in the model or residual systematics in the input data.  

\begin{figure*}
    \centering
     \includegraphics[width=1\linewidth]{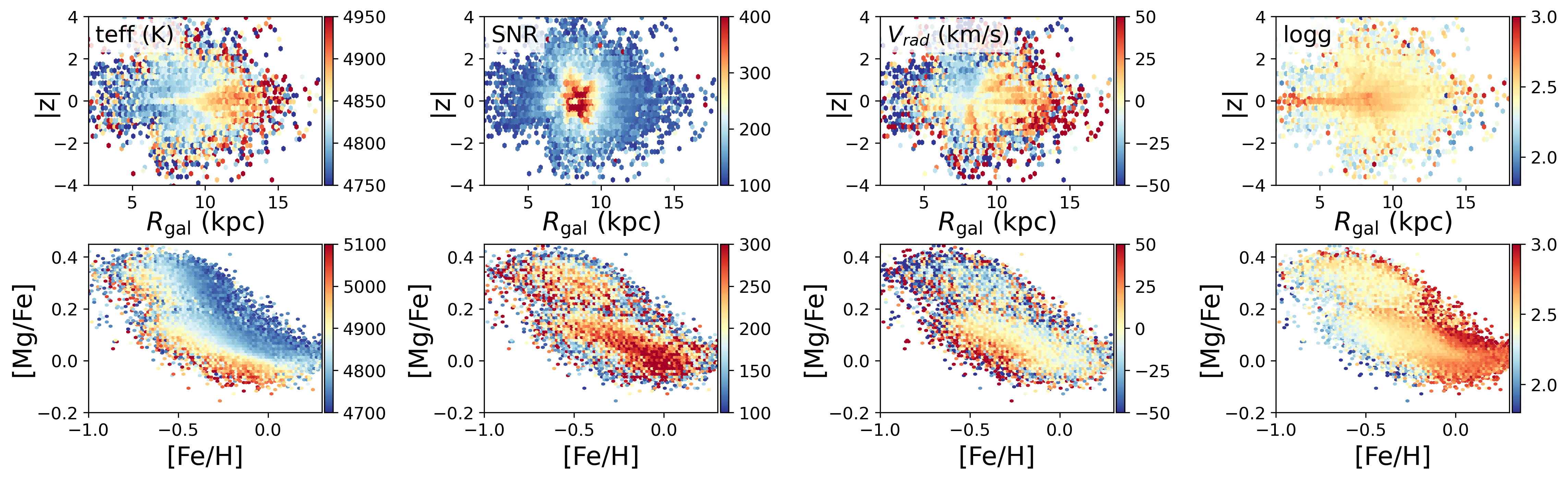}
    \caption{The maps of non-abundance parameters not included in the model including \teff, SNR, heliocentric velocity, $V_{\mathrm{rad}}$, \logg, which reveal some similarities to the residual structure shown in Figures \ref{fig:resid} and \ref{fig:resid2}.}
    \label{fig:resid3}
\end{figure*}

To investigate possible sources of these systematics in the input data, in Figure~\ref{fig:resid3} we map the median values of signal-to-noise ratio (SNR), effective temperature (\(T_{\mathrm{eff}}\)), surface gravity (\(\log g\)), and radial velocity in the same abundance and spatial domains. We find that these observational parameters also vary systematically across the sample, and in some cases display similar patterns to those seen in the element residuals.  

This similarity suggests that at least part of the residual structure may be due to correlated systematics in the observed spectra, for example, abundance measurement biases that depend on \(T_{\mathrm{eff}}\), \(\log g\), or SNR. Alternatively, residuals may trace real, physically motivated nucleosynthetic variation not captured by the adopted number of latent components. Disentangling these effects remains a challenge, but these residual patterns may ultimately provide a powerful diagnostic of both the model limitations and the quality of the input data.

\end{document}